# Effect of carbon-based nanoparticles on the ignition, combustion and flame characteristics of crude oil droplets

Gurjap Singh[a*], Mehdi Esmaeilpour[b], Albert Ratner[a]

[a] Department of Mechanical Engineering, The University of Iowa, Iowa City, IA 52242, USA

[b] College of Information Technology and Engineering, Marshall University, Huntington, WV 25755, USA

*Corresponding author: gurjap-singh@uiowa.edu*

**Abstract**

The use of in-situ burning (ISB) as a clean-up response in the event of an oil spill has generated controversy because of unburned hydrocarbons and products of incomplete combustion left behind on an ISB site. These substances threaten marine life, both in the ocean and on the ocean floor. Treating crude oil as a multicomponent liquid fuel, this manuscript investigates the effect of carbon-based nanomaterials, acetylene black (AB) and multi-walled carbon nanotube (MWNT), on the combustion and flame characteristics of crude sourced from the Bakken formation (ND, USA). Sub-millimeter droplets of colloidal suspensions of Bakken crude and nanomaterials at various particle loadings were burned, and the process was captured with CMOS and CCD cameras. The resulting images were post-processed to generate burning rate, ignition delay, total combustion time, and flame stand-off (FSR) ratio data for the various crude suspensions. A maximum combustion rate enhancement of 39.5% and 31.1% was observed at a particle loading of 0.5% w/w acetylene black nanoparticles and 0.5% w/w multi-walled carbon nanotubes, respectively. Generally, FSR for pure Bakken was noted as larger than for Bakken with nanoparticle additives. These results are expected to spur further investigations into the use of nanomaterials for ISB crude oil clean-ups.

**Abbreviations:**

AB: acetylene black
CCD: charge-coupled device
CMOS: complementary metal-oxide semiconductor
FSR: flame stand-off ratio
GC-MS: gas chromatography-mass spectrometry
ISB: in situ burning
MWNT: multi-walled carbon nanotube

**Highlights:**

- Carbon-based nanomaterials mixed into Bakken crude oil at various particle loadings
- Spherical sub-millimeter droplets burned using well-validated experimental setup
- Burning rate, ignition delay, total combustion time, flame standoff ratio compared
- Large increase in burning rate observed at small nanomaterial particle loadings
- Results expected to increase in-situ burning effectiveness as oil spill response

## 1. Introduction

Crude oil production has steadily increased since the 1940s due to worldwide population growth and increases in standards of living, and it is projected to keep increasing [1-2]. The shipping of crude oil has increased the risk of oil spills, which are defined as the release of liquid petroleum or crude oil (and their derivatives such as diesel) into the environment, especially the oceans and other marine environments. Many notorious oil spills on the high seas (such as the Deepwater Horizon incident) in the past have increased regulatory pressures on oil companies to take measures such as requiring double hulls on oil tankers [3] to decrease the number of these incidents. In addition to spills from oil tankers (which account for only 5% of total oil pollution), there are many other sources of oil spills, such as refinery terminals and oil wells [4]. Asia was the largest source of major oil spills from 1960 to 2010, with almost 3.4 million tons of oil released into the environment [5]. In the US, there are almost 25 spills per day into navigable waters [4].

Crude oil spill control and clean-up is a challenging topic with major implications for marine pollution management and environmental conservation. Large oil spills and the subsequent large-scale spill responses attract much public attention: the 1969 Santa Barbara oil spill that released 14,300 tons of crude oil is credited for providing the impetus for establishment of the US Environmental Protection Agency [3,6]. The aforementioned Deepwater Horizon incident released over 700,000 tons of crude oil into the oceans [7], which caused health issues for 40% of Gulf Coast residents, caused 5.5 billion dollars in damage to fishing and tourism industry, and polluted 692 km of marsh shoreline [5].

Because of the severe economic, human health, and environmental impacts of marine oil spills, many methods have been explored to develop spill response strategies. Their effectiveness depends on the size of the spill and crude oil composition. For a small oil spill, sorbents such as polyurethane pads can be used to soak up the crude oil on the surface [8], or solidifying agents can be used to constrain the spilled oil and recover it as a solid [9]. For larger oil spills, the oil is constrained using booms and can then be skimmed off using mechanical devices [10]. Chemical dispersants can be added to the oil spill to promote its emulsification into the ocean before it travels to the shore, which also promotes its biodegradation [11]. Bioremediation by using nitrogen-based fertilizers can also be used to promote biodegradation of the crude oil by microbes [12]. In situ burning (ISB) of crude oils, one of the oldest countermeasures for oil spills, remains in use today as well. All these methods have their drawbacks. For example, concerns have been raised over the aquatic toxicity of solidifiers and chemical dispersants. In situ burning, although desirable because of its ease of use in remote offshore areas, can leave significant amounts of unburned hydrocarbons in the water and releases a large amount of particulate matter into the air, contributing to marine and air pollution. Its effectiveness also depends on the type of crude being used, with heavy, emulsified crudes and weathered crudes having medium to low ISB burnability [5].

Systematic studies of crude oil combustion are rare in scientific literature and mostly related to pool fires. Previously, crude oil droplet combustion has been explored in the work of Singh *et al*. [13], where a comparison of the combustion and sooting properties of different US crudes (Bakken, Colorado, Pennsylvania, and Texas) was presented. It was found that crudes from Bakken, Colorado, and Pennsylvania burned at comparable rates, but Texas crude oil burned predominantly with violent microexplosions. Similarly treating crude oil as a multicomponent liquid fuel, this work develops a method to improve the combustion efficiency of ISB by using carbon-based nanoparticles to increase the heat conductivity and radiation absorption of the crude oil. Note that ISB effectiveness has been reported to be positively dependent on radiation absorption [14]. Combustion properties of single spherical fuel droplets have been presented, which are expected to aid eventual mathematical modeling of the combustion process. The results also support the development of a mechanical device that will burn the crude skimmed off the spill surface in a droplet spray regime after mixing it with nanoparticles, which will lead to better burnability for heavy, emulsified, and weathered crudes, better combustion efficiency, and less particulate emissions than pool fires.

The well-established experimental method used for this work has previously been used to characterize various combustion properties for liquid fuels, such as burning or combustion rate [13, 15-19], ignition delay [13, 18], and total combustion time [13, 18]. This benchtop method uses only a small amount of sample, which is important because of the high cost of the nanomaterials involved, especially multi-walled carbon nanotubes (MWNT).

There are many challenges associated with the characterization of a multicomponent liquid fuel such as Bakken crude oil. It is a light and sweet variety of crude oil, composed mainly of low-to-medium boiling hydrocarbon fractions (see Appendix A). The presence of components that boil at different temperatures causes the burning droplet to shatter into fragments during the combustion process in an event called a microexplosion. Despite microexplosions, a significant part of the combustion regime follows the “$d^2$-law”. This part of the combustion regime across different fuels was used to calculate and compare combustion rates.

## 2. Methods and materials

The method used for this work has previously been detailed by Singh *et al.* [13, 18] and is based on the original work of Avedisian and Callahan [20] and Bae and Avedisian [21]. A brief explanation is as follows: three 16 µm SiC fibers are fixed on six poles arranged in a circle such that the fibers cross at the center (see **Figure 1**). A sub-millimeter fuel droplet is placed using a micro-syringe and ignited symmetrically using two hot wire loops on either side. As the droplet heats up and burns, the process is recorded using two high-speed cameras. One of the cameras is a black and white charge-coupled device (CCD) IDT X-StreamVision XS-3 (IDT Vision, Pasadena, CA, USA) operated at 1000 frames/second and fitted with a 105 mm lens (Nikon AF Micro-Nikkor-F/2.8, Tokyo, Japan), with backlighting provided by a single bright white LED at 3.3V. The other high-speed camera is a complementary metal-oxide semiconductor (CMOS) Casio EXILIM Pro EX-F1 (Casio, Tokyo, Japan) operating at 600 frames/second and recording the droplet combustion scene through a magnifying concave mirror (4.0” diameter, 9.0” focal length).

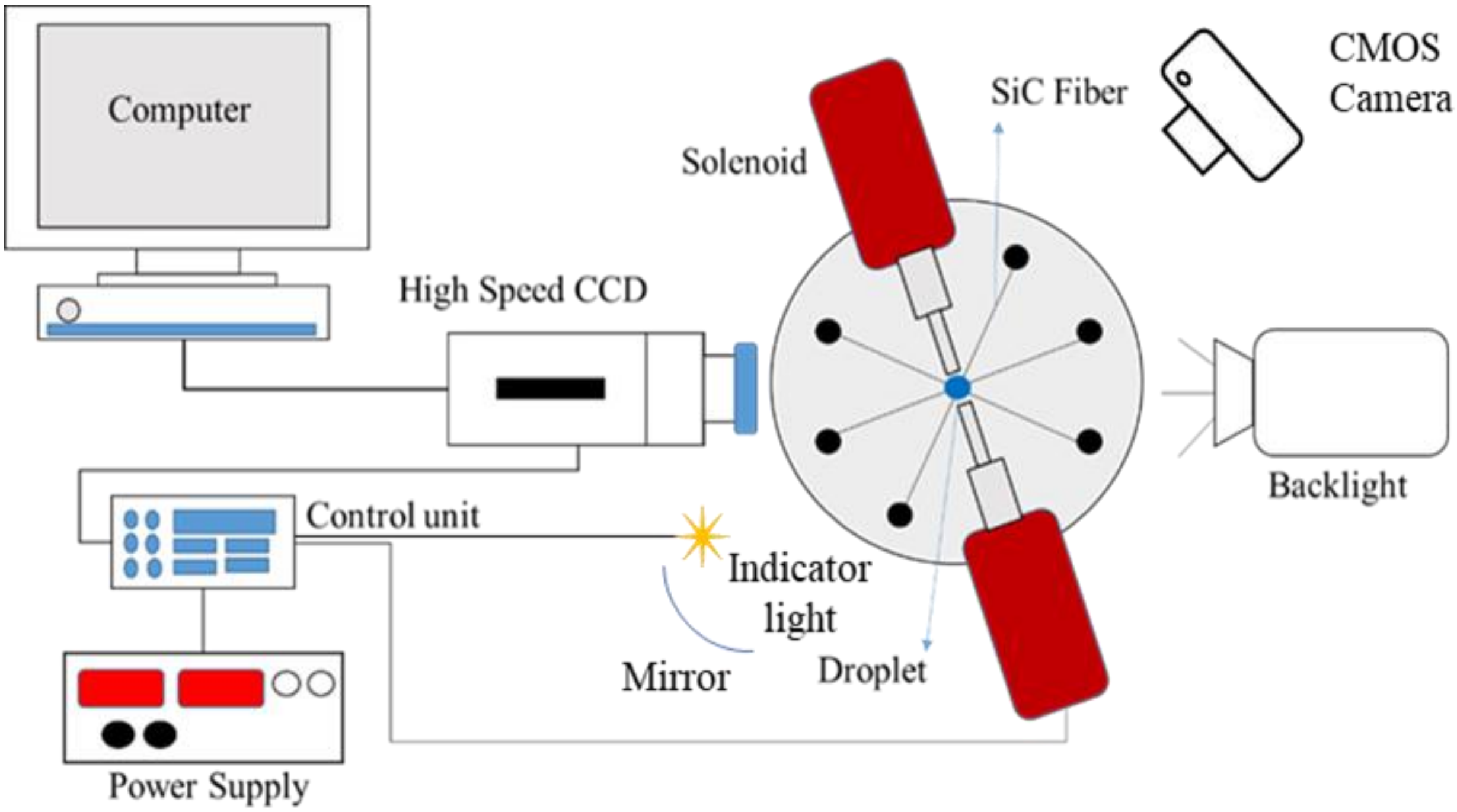

**Figure 1.** Experimental setup schematic showing the location of the fuel droplet and various components [13].

The CCD camera generates 8-bit grayscale image data with a resolution of 948 x 592 pixels. It is post-processed using ImageJ, an open-source digital image processing software [22-24] to generate a time series for the droplet area as it shrinks because of combustion. This data is used to find combustion rate and flame stand-off ratio. The CMOS camera generates color video footage that is post-processed using MATLAB ® (MathWorks, Natick, MA, USA) to generate time-series 8-bit/channel RGB frames with a resolution of 432 x 192 pixels. This data is used to determine the ignition delay and total combustion time.

The initial droplet size $d_0$ is sub-millimeter because larger droplets cannot be supported by the SiC support fibers, but must exceed the criteria described by Avedisian and Jackson [25] to keep $d_0/d_{fiber} > 13$ to match burning rates to unsupported droplets. Due to the small size of the supporting fibers, the droplet remains almost perfectly spherical during initial set-up and during most of the combustion process, other than the violent microexplosion regime.

The base liquid fuel used in this work is Bakken crude oil acquired from the United States Northern Great Plains oil production region (North Dakota Bakken formation). Although crude oil properties (even from the same oilfield) may vary significantly, the typical specific gravity of Bakken crude is 0.815, and its initial boiling point is 21 ºC [26]. Appendix A contains gas chromatography-mass spectrometry (GC-MS) data for the sample used in this work, showing a multicomponent liquid that has a significant amount of low and medium boiling constituents.

Two types of carbon-based nanomaterials including acetylene black (AB) and multi-walled carbon nanotube (MWNT) are used in this work. Acetylene black (AB) is a well-characterized electrically conductive material that is used as a charge collector in chemical batteries [27-29]. The AB used for this experimental work was obtained from Alfa Aesar (Ward Hill, MA, USA). It is a 100% compressed, 99.9% purity dark black solid powder with specific surface area 75 $m^2/g$, bulk density 170-230 g/L, and mean particle size 25-45 nm. The MWNT used for this experimental work were obtained from Nanostructures & Amorphous Materials (Katy, TX, USA). These are short variety nanotubes with >95% purity, outside diameter 8-15 nm, inside diameter 3-5 nm, length 0.5-2 µm, specific surface area 233 $m^2/g$, and bulk density 360-420 g/L.

Colloidal suspensions of 0.5%, 1%, 2%, and 3% w/w were prepared from the nanomaterials by mixing them with Bakken crude oil in 20 mL disposable scintillation glass vials (Fisher Scientific, Hampton, NH, USA). An ultrasonic disruptor (Biologics 3000 MP) with a 3/16" probe was used to mix the sample for 5 minutes. To minimize heat generation, the ultrasonicator was programmed to operate on pulses that were 4 s long with 4 s between them. Note that both the AB and MWNT colloidal suspensions prepared using this method have previously been found to form very stable suspensions in hydrocarbon-based liquid fuels [30].

## 3. Results and discussion

Many physical phenomena play a role in determining properties such as ignition delay, combustion rate, total combustion time, and flame standoff ratio of crude oil droplets. Addition of a solid dispersed phase (nanomaterials) leads to the decrease in vapor pressure of the continuous (liquid) phase, but in this case the solid dispersed phase forms a highly interconnected latticework of individual nanoparticles. Through radiometric temperature measurements, addition of these nanomaterials has been observed to raise the droplet temperature for multicomponent fuels during combustion, due to radiation absorption and increase in bulk heat conductivity [18]. Furthermore, no significant differentiation has been observed between the droplet temperature at different particle loadings, so long as nanomaterials are present. This points to an upper limit to the increase in radiation absorption and heat conductivity through addition of nanomaterials. In addition, many long-chain hydrocarbon-based crude oil constituents are expected to thermally decompose into lighter, smaller components at high temperatures as the combustion process progresses. As temperatures are higher in droplets with nanomaterials, more decomposition and lighter component evolution can be expected compared to pure crude droplets, leading to more flame speeds. However, the presence of nanomaterials on droplet surface hinders liquid evaporation, which leads to very non-linear and complex results at different particle loadings for different nanomaterials.

As reported previously by Singh *et al*. [13], most of the combustion process of the pure Bakken crude oil is marked by the presence of microexplosions because of its highly multicomponent nature (see **Figure 2 a-e**). For Bakken crude, four distinct zones are present. Zone I is ignition delay, for the duration of which the droplet expands volumetrically from initial diameter $d_0$ as it is heated. The end of Zone I is marked by a decrease in normalized droplet area, when the fuel in the droplet starts to evaporate and burn. This is the beginning of Zone II, which is marked by steady combustion. Because of high heat in the droplet, the various medium-to-high boiling fractions present in the fuel start to bubble and cause microexplosions, in a mechanism that has been detailed in the previous work of Singh *et al*. [13]. The zone with violent microexplosions is Zone III. Eventually, high-intensity microexplosions give way to low-intensity microexplosions with steady combustion, which is Zone IV.

When AB or MWNT nanomaterials are added to the crude oil, another combustion zone is added after Zone IV, where the nanomaterial residue itself is observed to burn as a char. This is termed Zone V (see **Figure 2 b-e**). The more nanomaterial added, the more char residue left and the longer Zone V lasts. As nanomaterials are added, visual inspection shows that the microexplosions in Zone III decrease in duration and intensity. Acetylene black is more effective at decreasing microexplosions than MWNT, as shown in **Figure 2** d (compare with **Figure 2** a) and as previously reported by Singh *et al*. [18]. It is also observed that more AB leads to lesser microexplosions.

The following sections discuss combustion properties such as combustion or burning rate, ignition delay, total combustion time, and flame stand-off ratio.

(a)

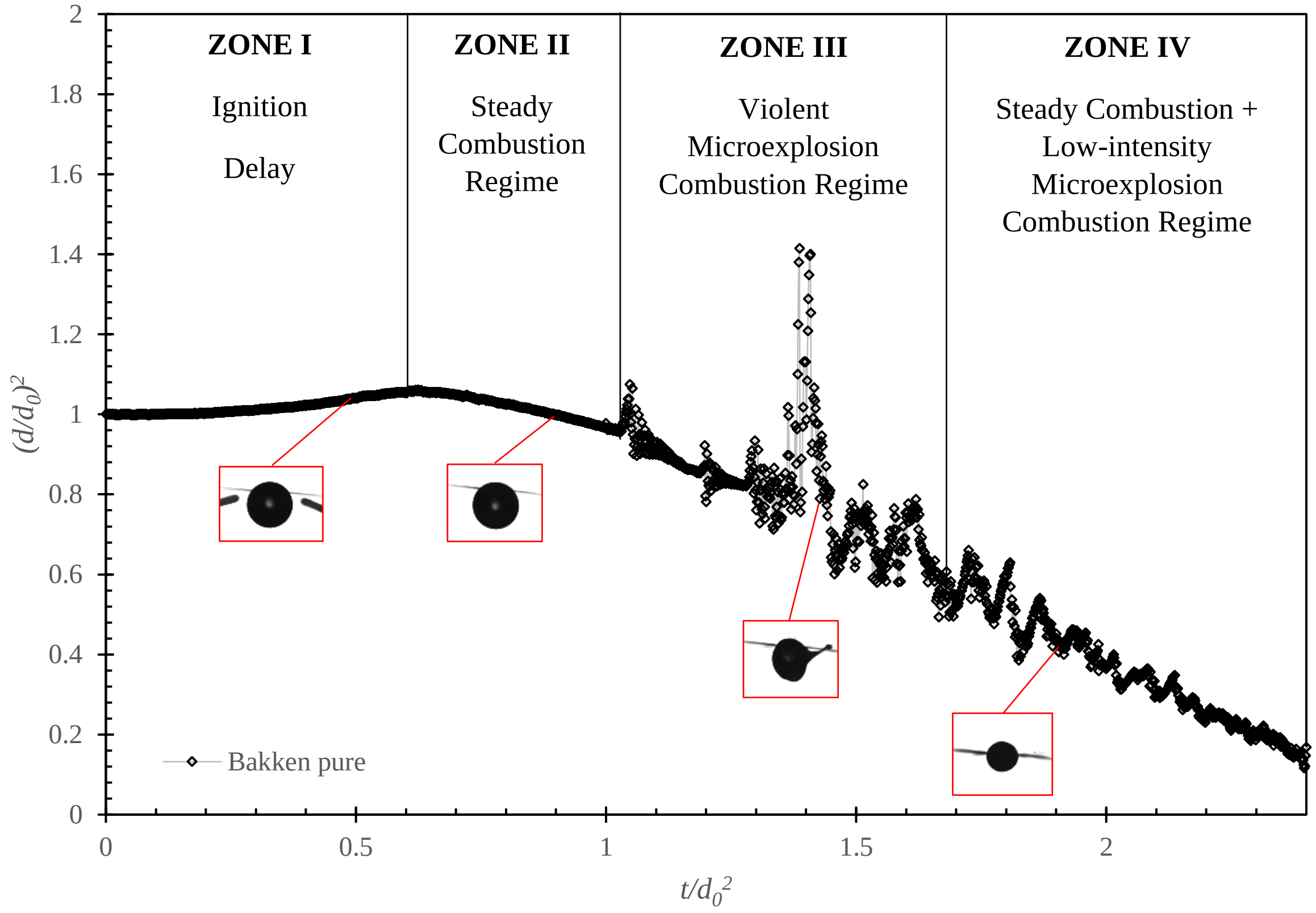

(b)

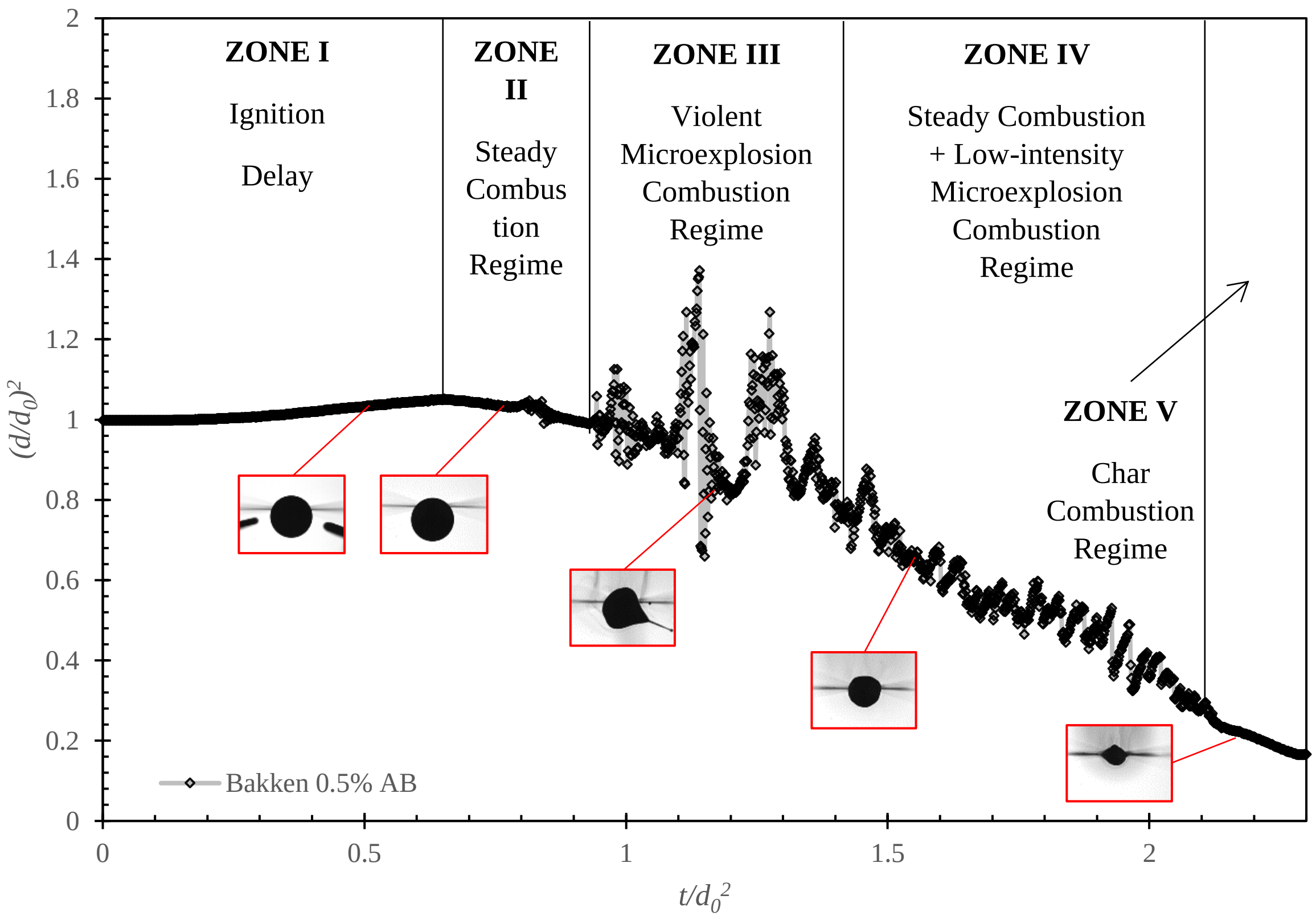

(c)

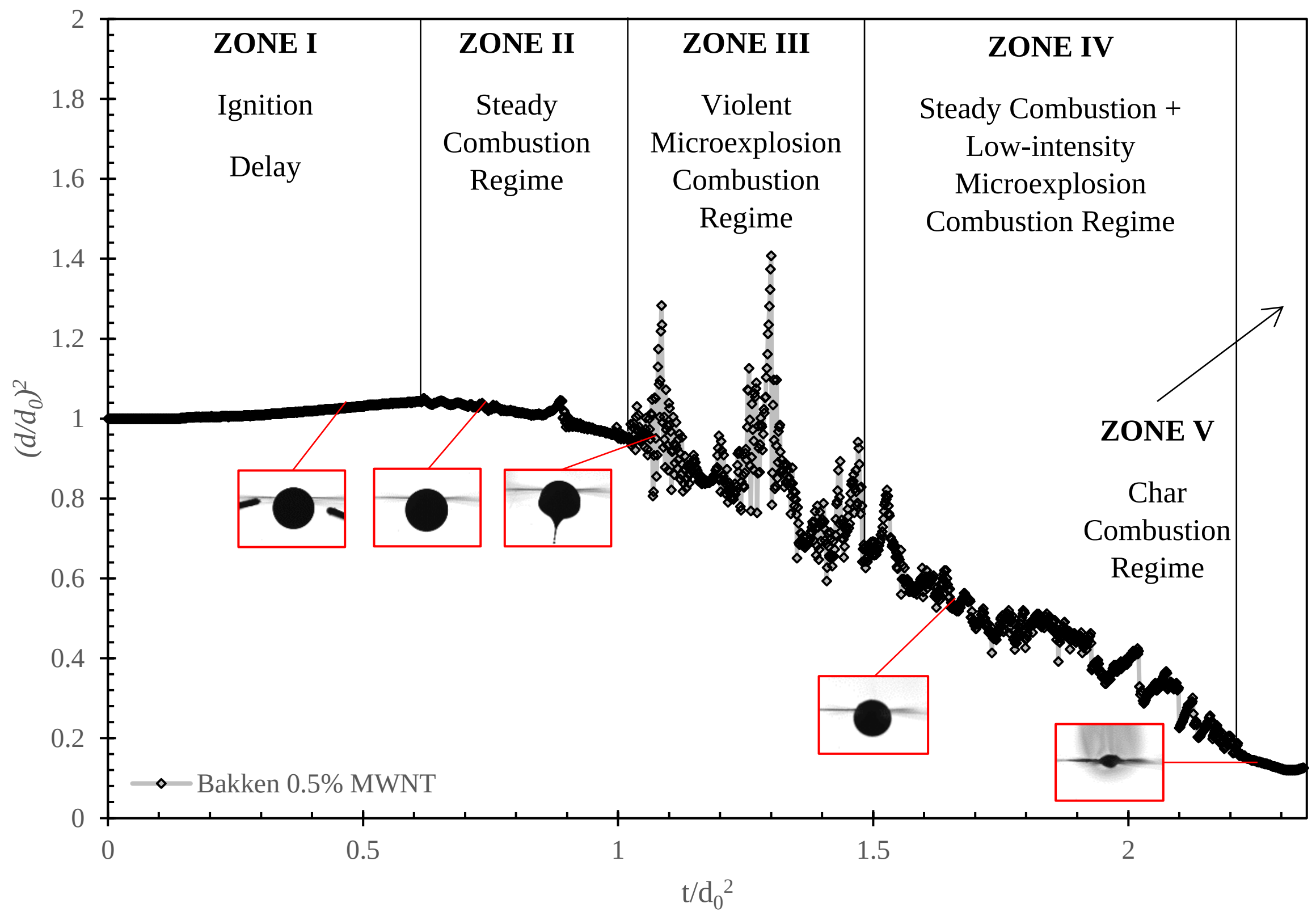

(d)

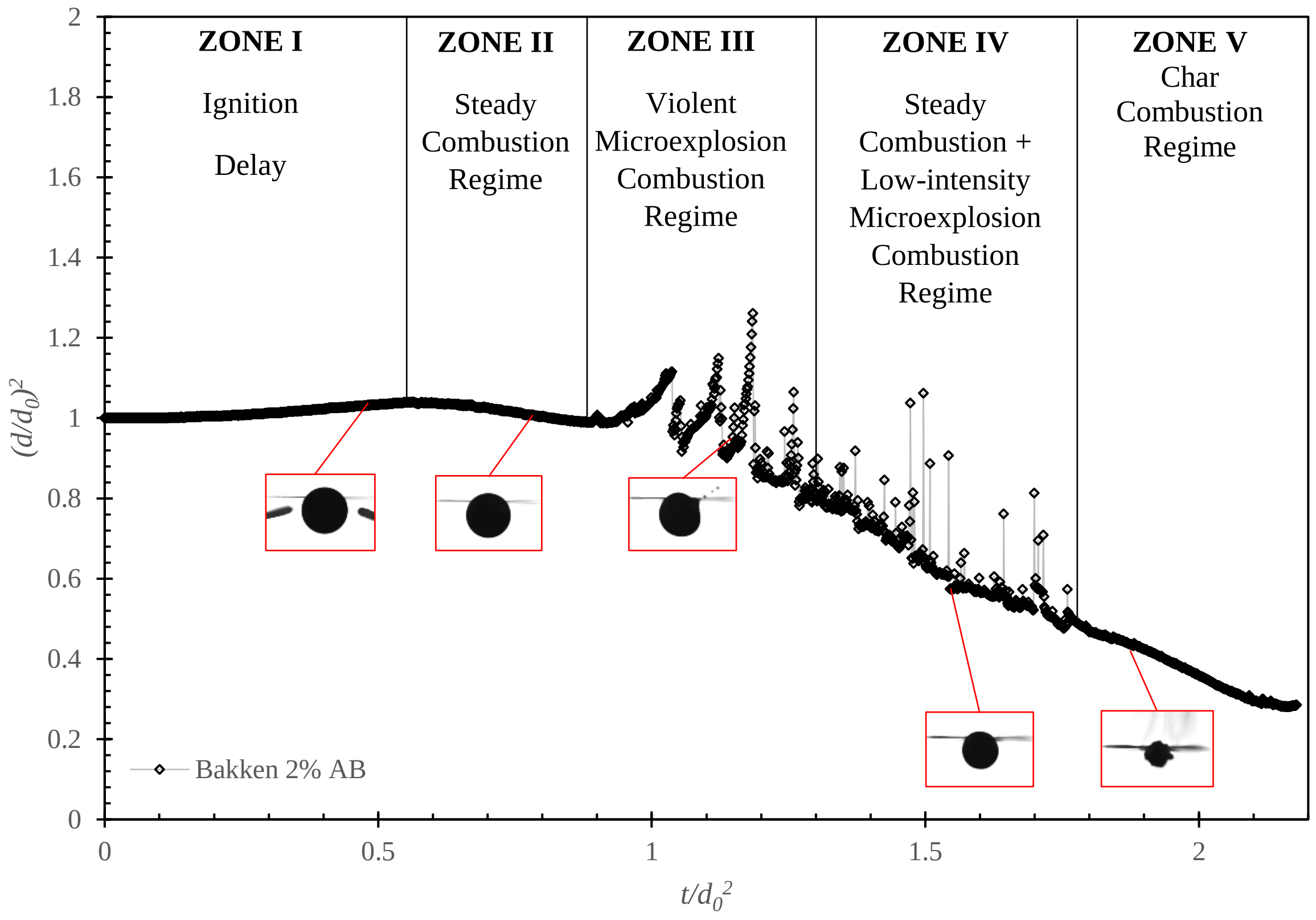

(e)

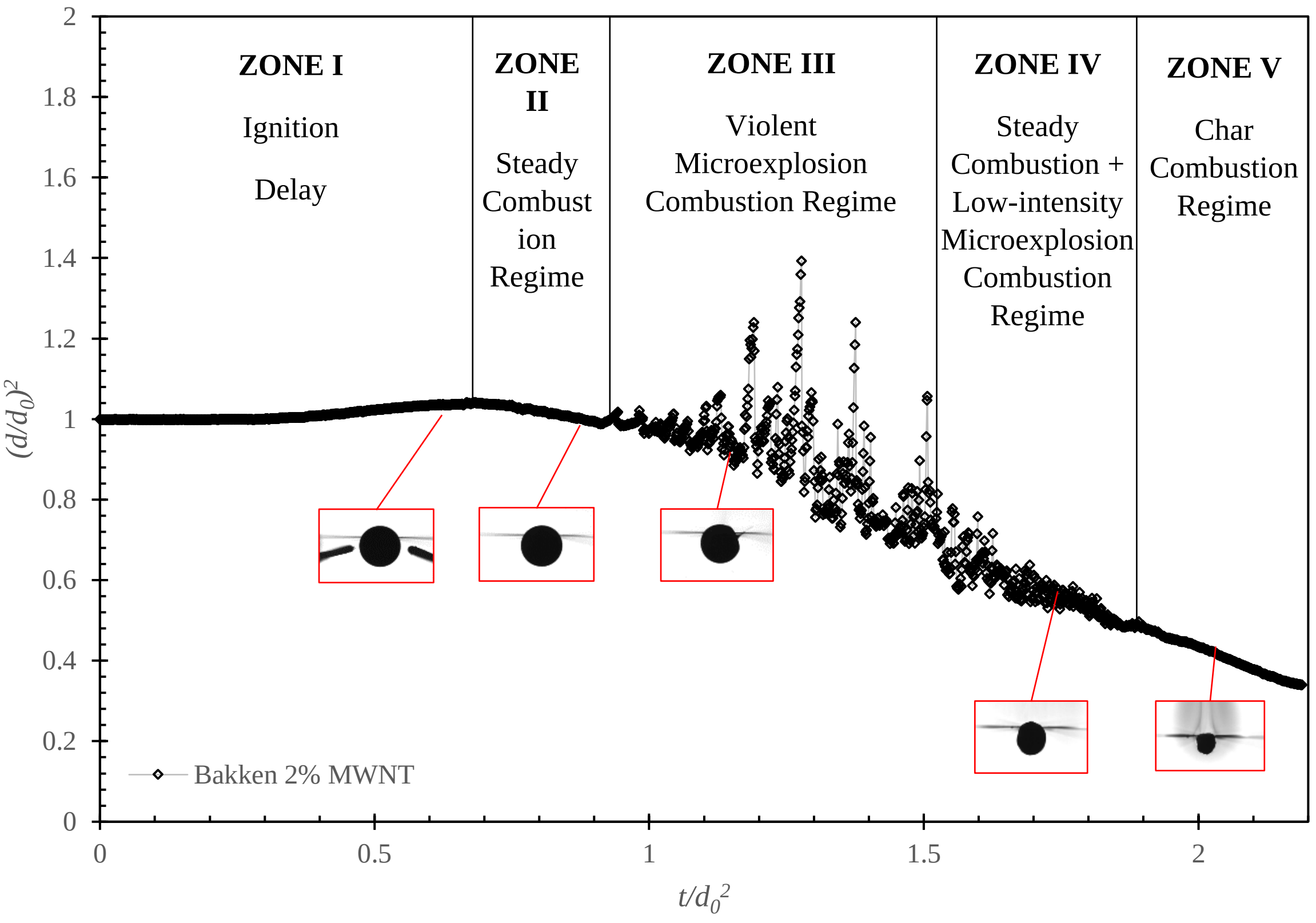

**Figure 2.** Evolution of normalized diameter $(d/d_0)^2$ with normalized time $t/d_0^2$ for (a) pure Bakken crude oil, (db) Bakken crude oil with 0.5% AB w/w particle loading, (dc) Bakken crude oil with 0.5% MWNT w/w particle loading, (d) Bakken crude oil with 2% AB w/w particle loading, (e) Bakken crude oil with 2% w/w MWNT particle loading. Snapshots taken from CCD camera representing various zones are included.

*3.1 Burning rate of fuel droplets*

As mentioned earlier, the fuel droplets being investigated went through many distinct combustion zones. Zone III (the violent microexplosion combustion regime) does not show a meaningful combustion trend, whereas Zone IV (the steady combustion + low intensity microexplosion regime) shows a much clearer and more meaningful trend. It also comprises a significant portion of the total combustion regime in all experiments conducted. Zone IV was therefore selected for burning rate comparison across the board. Zone V, or char combustion, is not present for pure Bakken droplets; therefore, it was not considered for this analysis. **Figure 3a-d** and **Figure 4a-d** show normalized droplet area comparison for Zone IV between Bakken crude and Bakken crude at various particle loadings for single droplets of comparable sizes.

The fuel droplet shrinks as it burns. The data from post-processing yields the change in droplet diameter $d(t)$ with time $t$ at 1000 frames/second. This data is reduced using the moving averages method to plot the combustion trend. The combustion trend in Zone IV follows the well-known $d^2$ law [31-32] :

$$\frac{d(t)^2}{d_0^2} = 1 - k(\frac{t}{d_0^2}) \qquad (1)$$

where $d_0$ is the initial droplet diameter and $k$ is the combustion or burning rate. The higher the $k$, the faster the duel droplet burns.

(a)

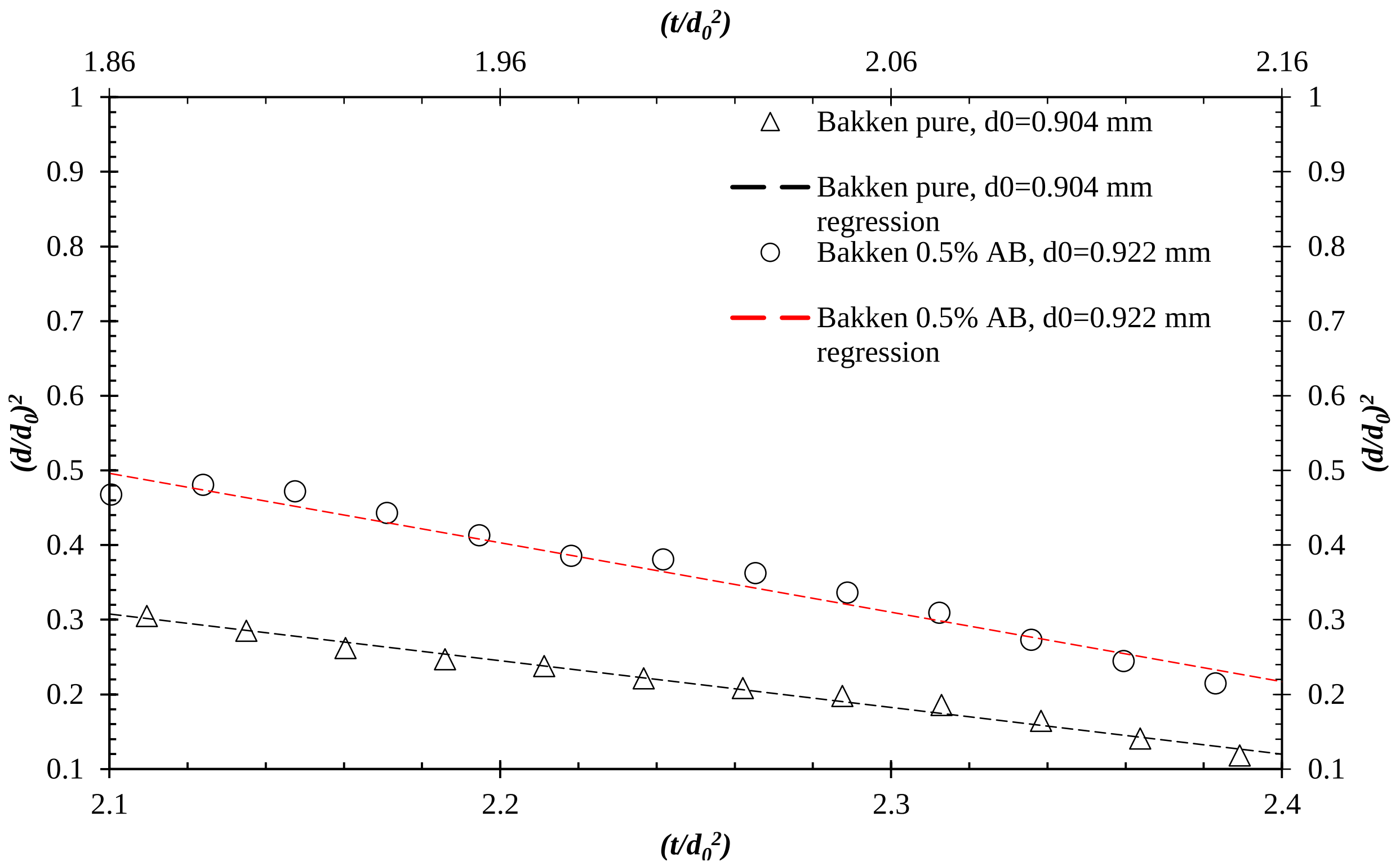

(b)

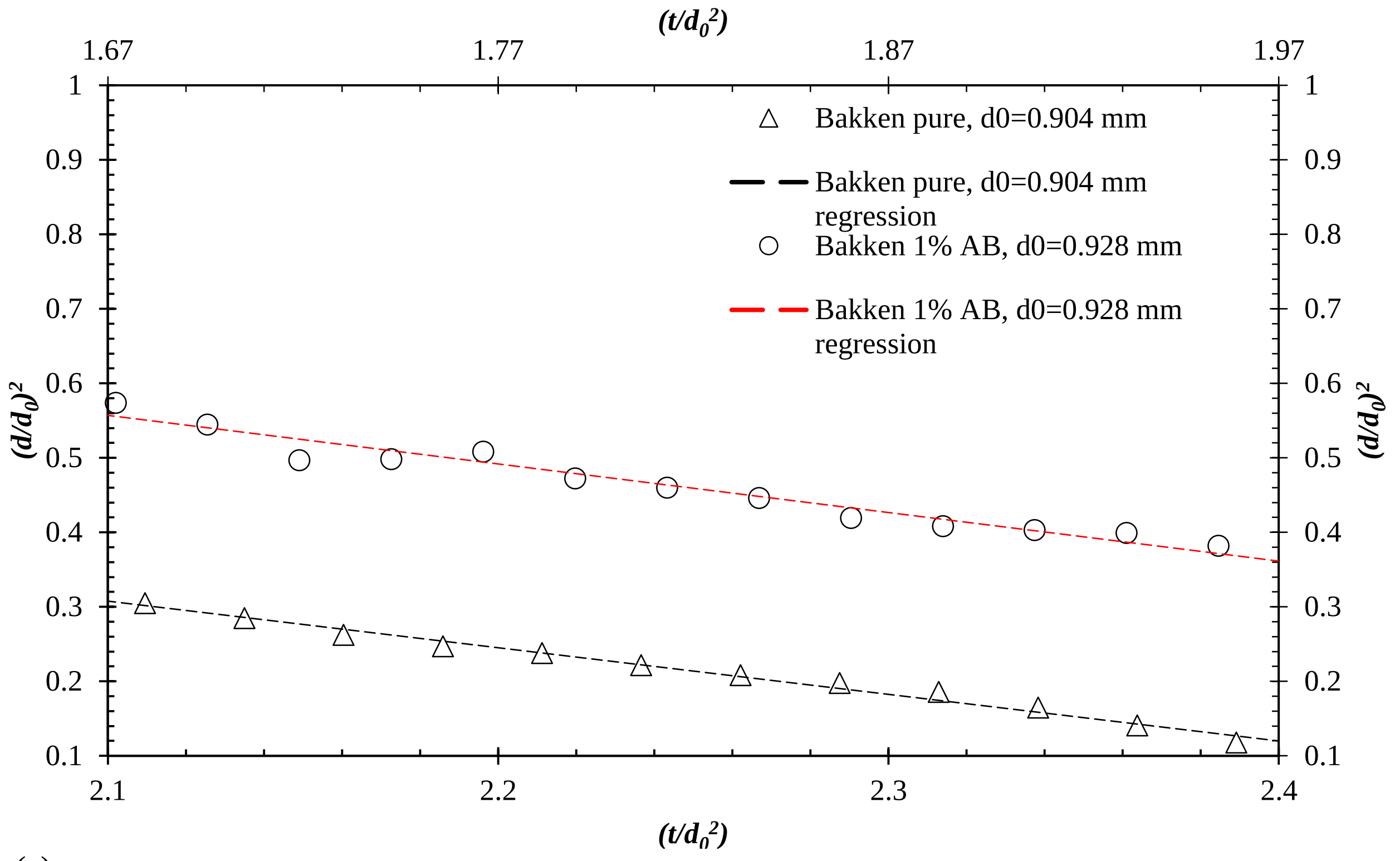

(c)

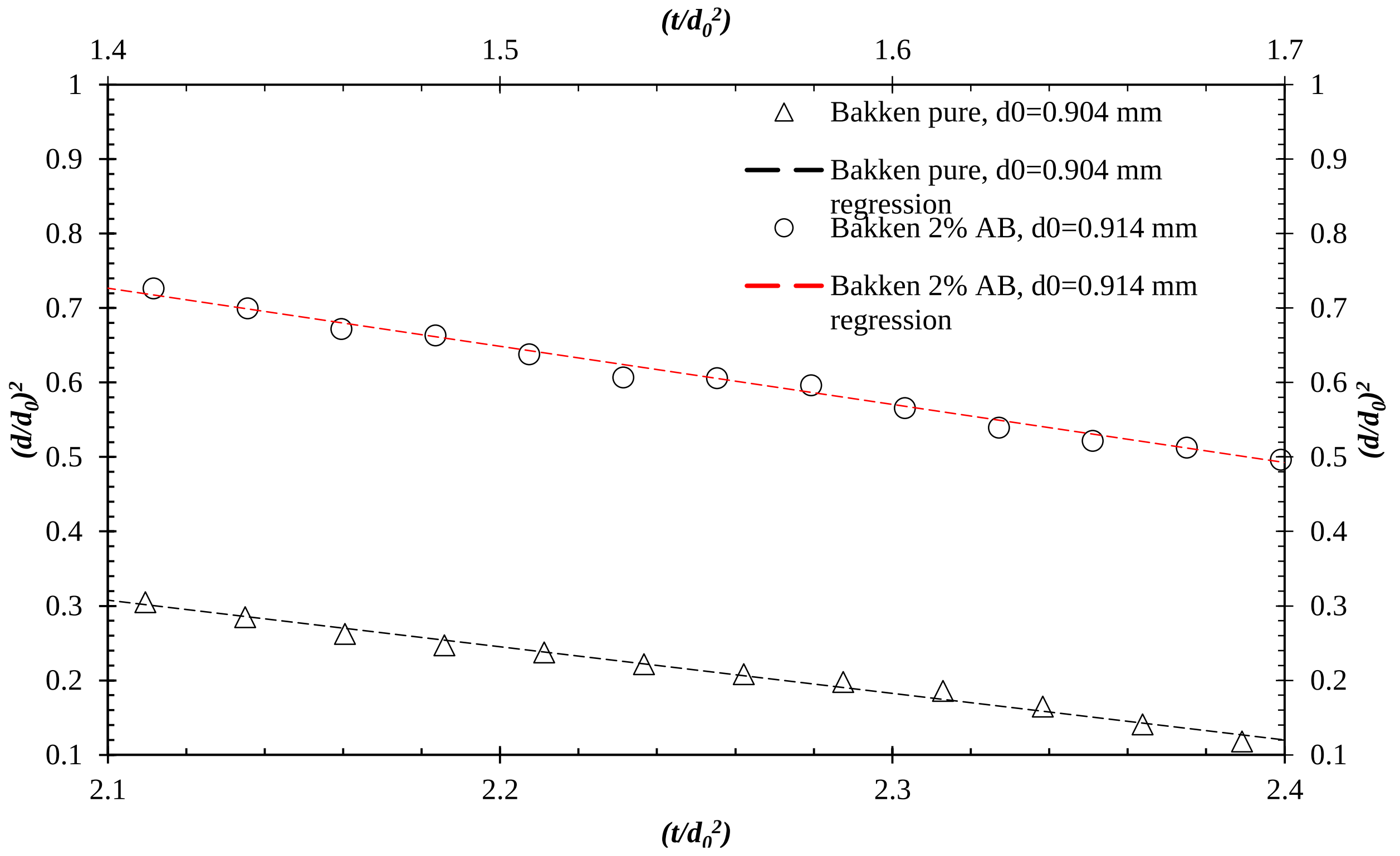

(d)

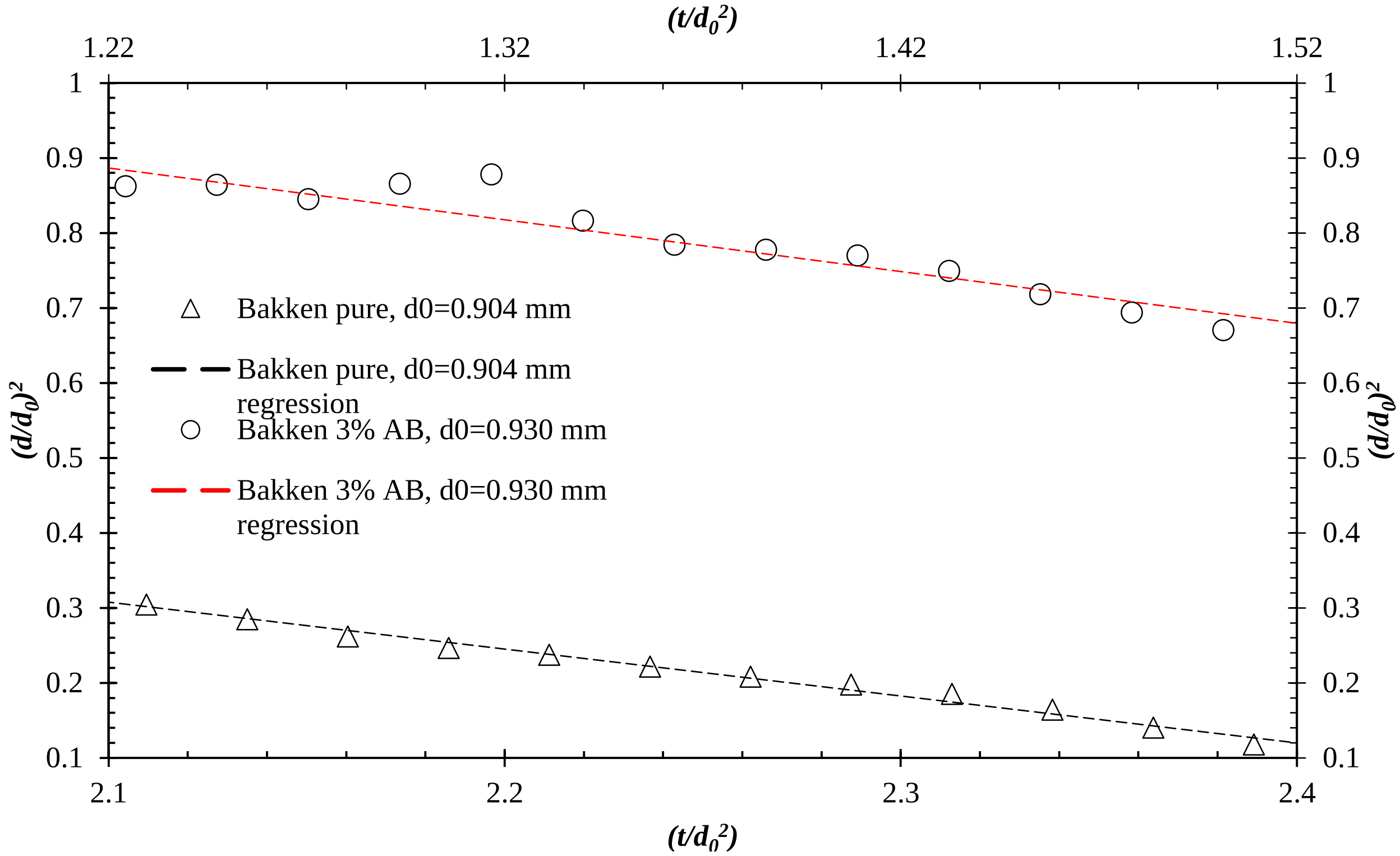

**Figure 3**. Zone IV comparison between Bakken crude and Bakken crude at different AB particle loadings. For single droplets of comparable sizes, moving average data comparison for normalized droplet area and trendlines are shown.

(a)

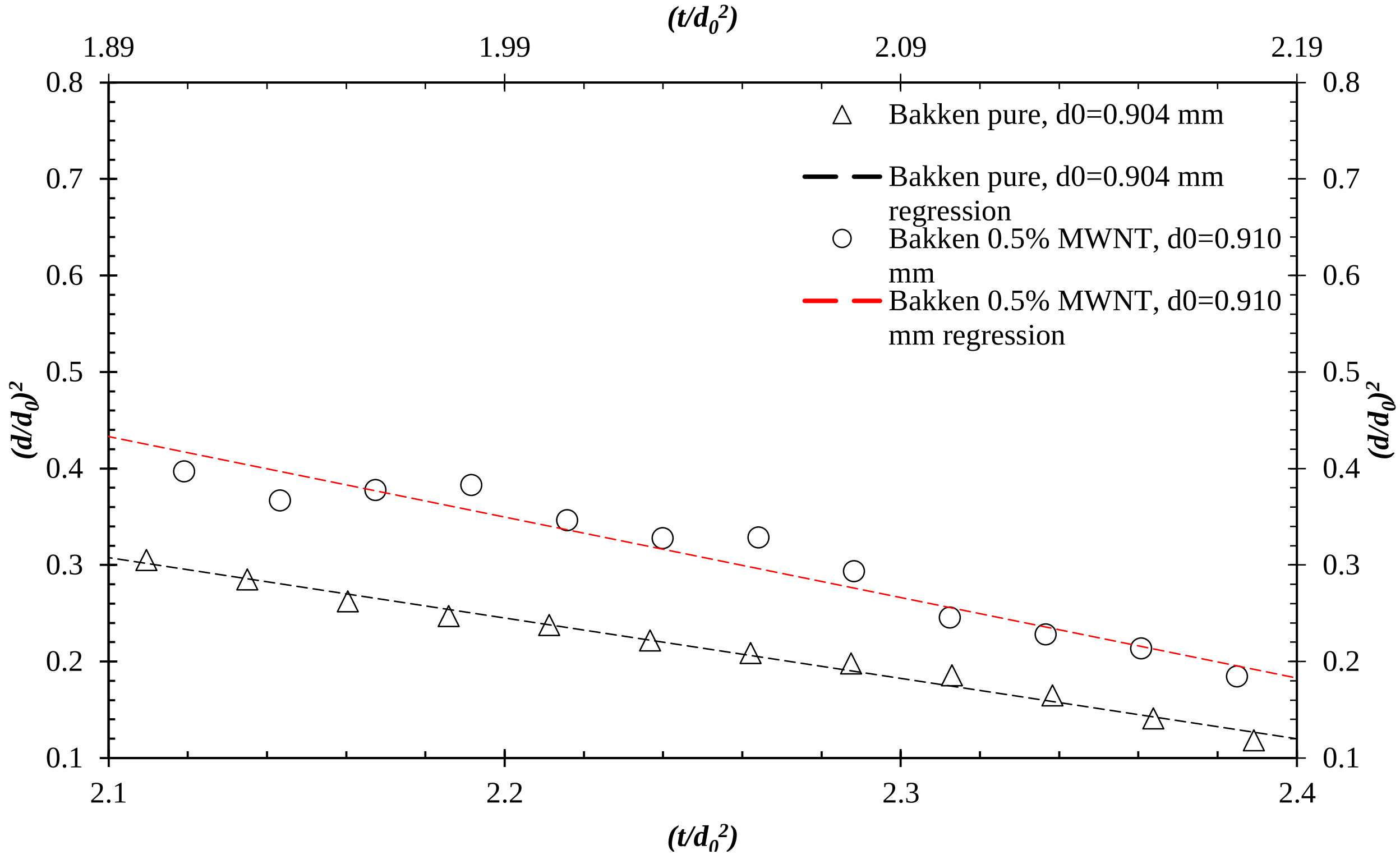

(b)

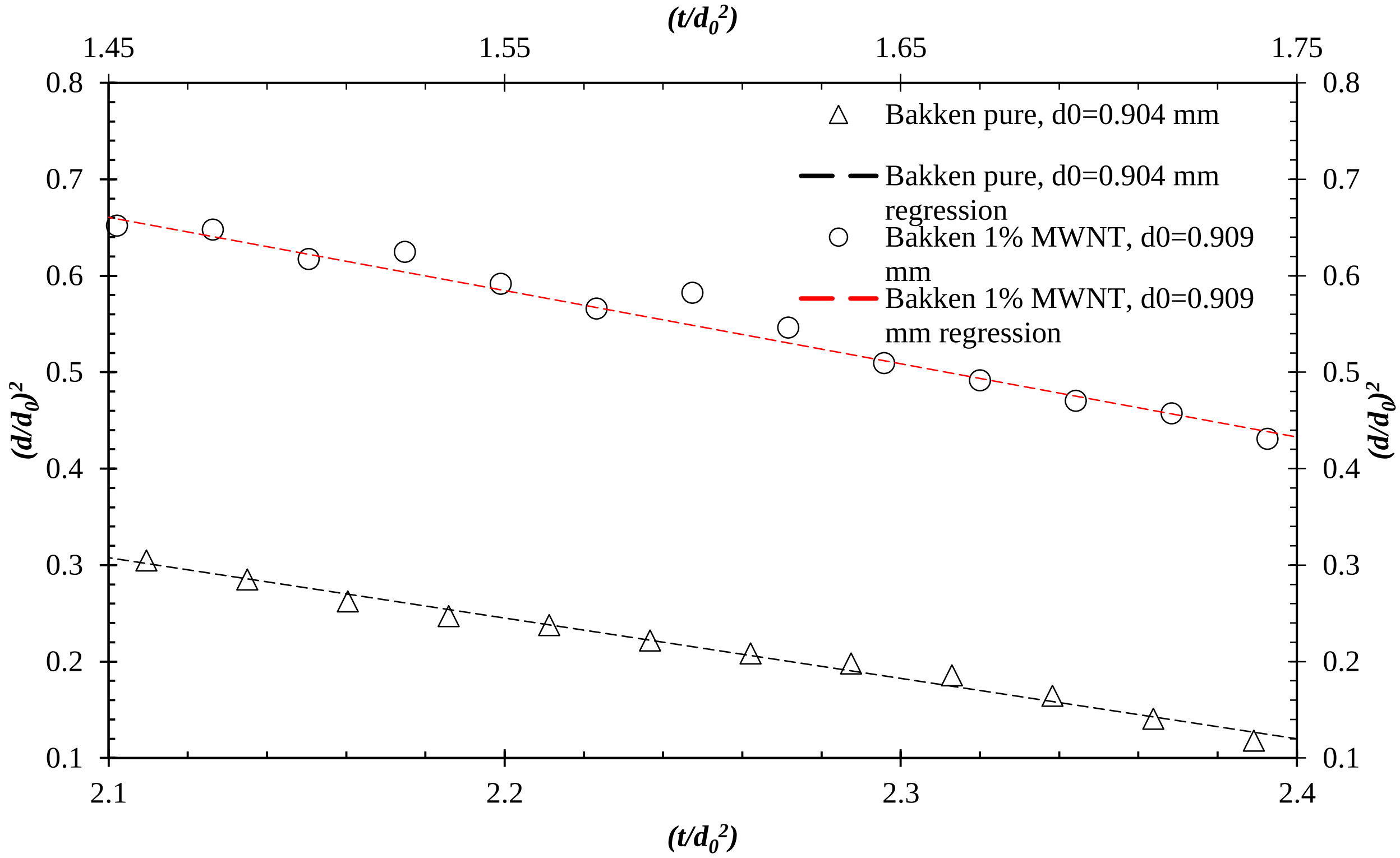

(c)

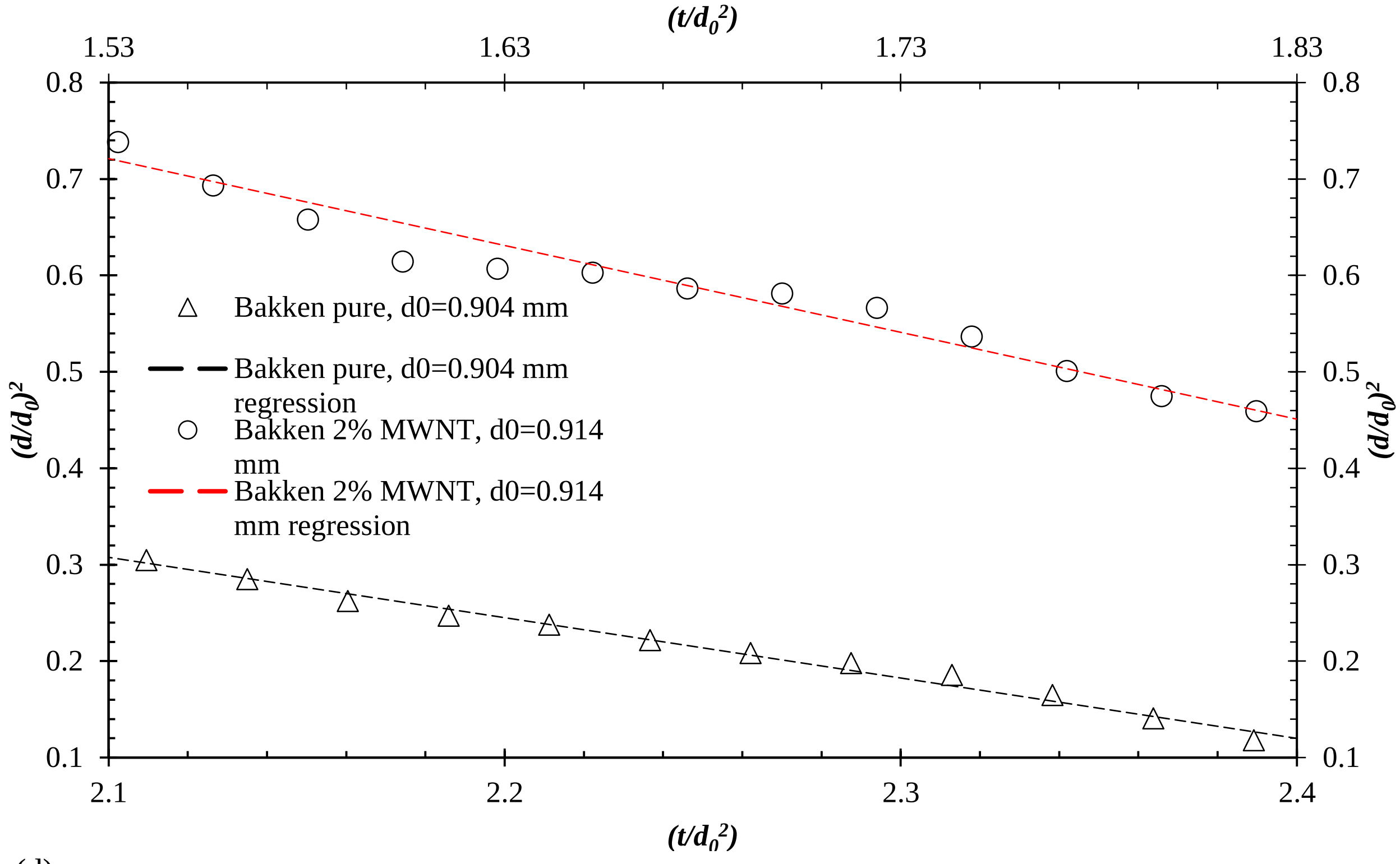

(d)

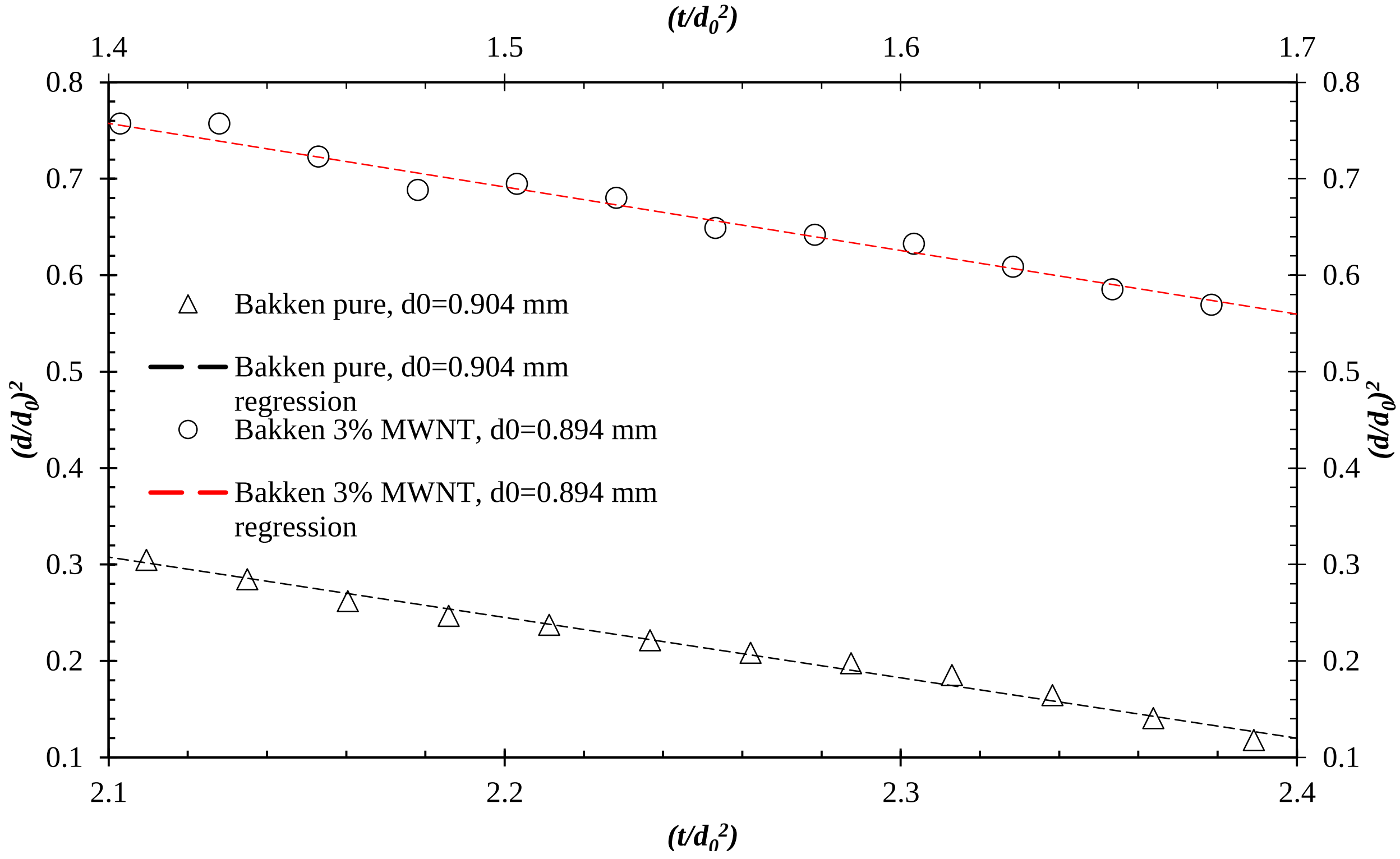

**Figure 4.** Zone IV comparison between Bakken crude and Bakken crude at different MWNT particle loadings. For single droplets of comparable sizes, moving average data comparison for normalized droplet area and trendlines are shown.

**Figure 5** shows the average burning rate comparison of Bakken crude at different nanomaterial loadings. Five experiments were conducted to determine the average ignition delay for all fuels. The error bars show the standard deviations of each set of experiments. As shown in this figure, a general increase in combustion rates is observed as both AB and MWNT are added.

(a)

(b)

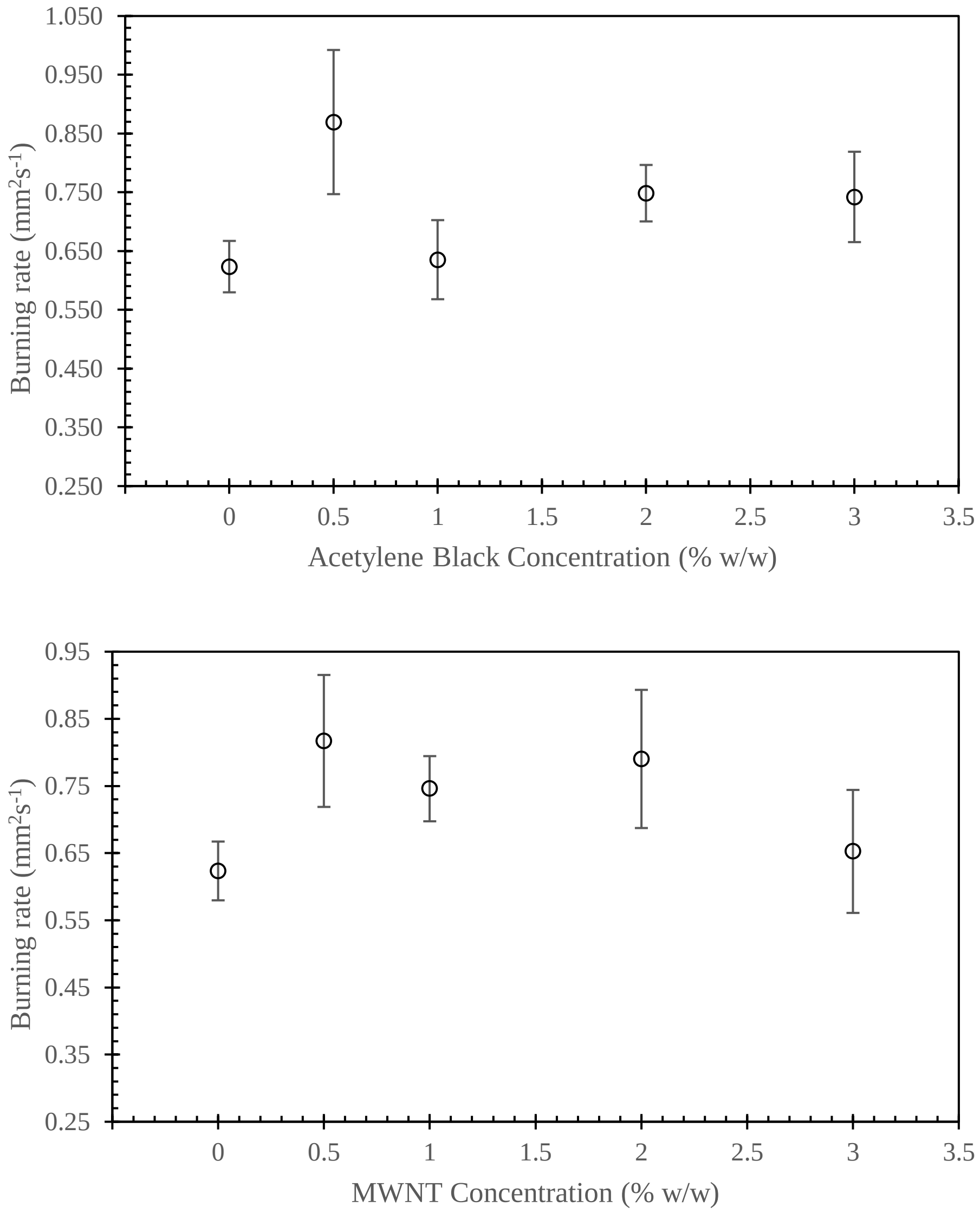

**Figure 5.** Comparison of the burning rates of Bakken crude with different mass concentrations of (a) AB and (b) MWNT.

Many physical and chemical properties determine the combustion rate of a burning liquid droplet, such as heat conductivity, radiation absorption, vapor pressure, and chemical composition. The greater the heat conductivity and radiation absorption of the droplet, the faster the combustion rate. The higher the vapor pressure, greater the vapor release rate and the higher the combustion rate. The addition of nanoparticles decreases vapor pressure but increases heat conductivity and radiation absorption, so complex results are possible from varying the particle loading in the liquid fuel.

**Figure 6** shows the change in average burning rate for Bakken crude at different particle loadings of AB and MWNT. The most significant change is seen at 0.5% w/w particle loading for both AB and MWNT, where 39.5% and 31.1% combustion rate enhancements were observed for these nanomaterials, respectively. Acetylene black shows an almost 20% combustion rate enhancement at 2% and 3% particle loading, which is also significant, and MWNT shows an enhancement of 19.6% at 1% particle loading and 26.6% at 2% particle loading.

(a)

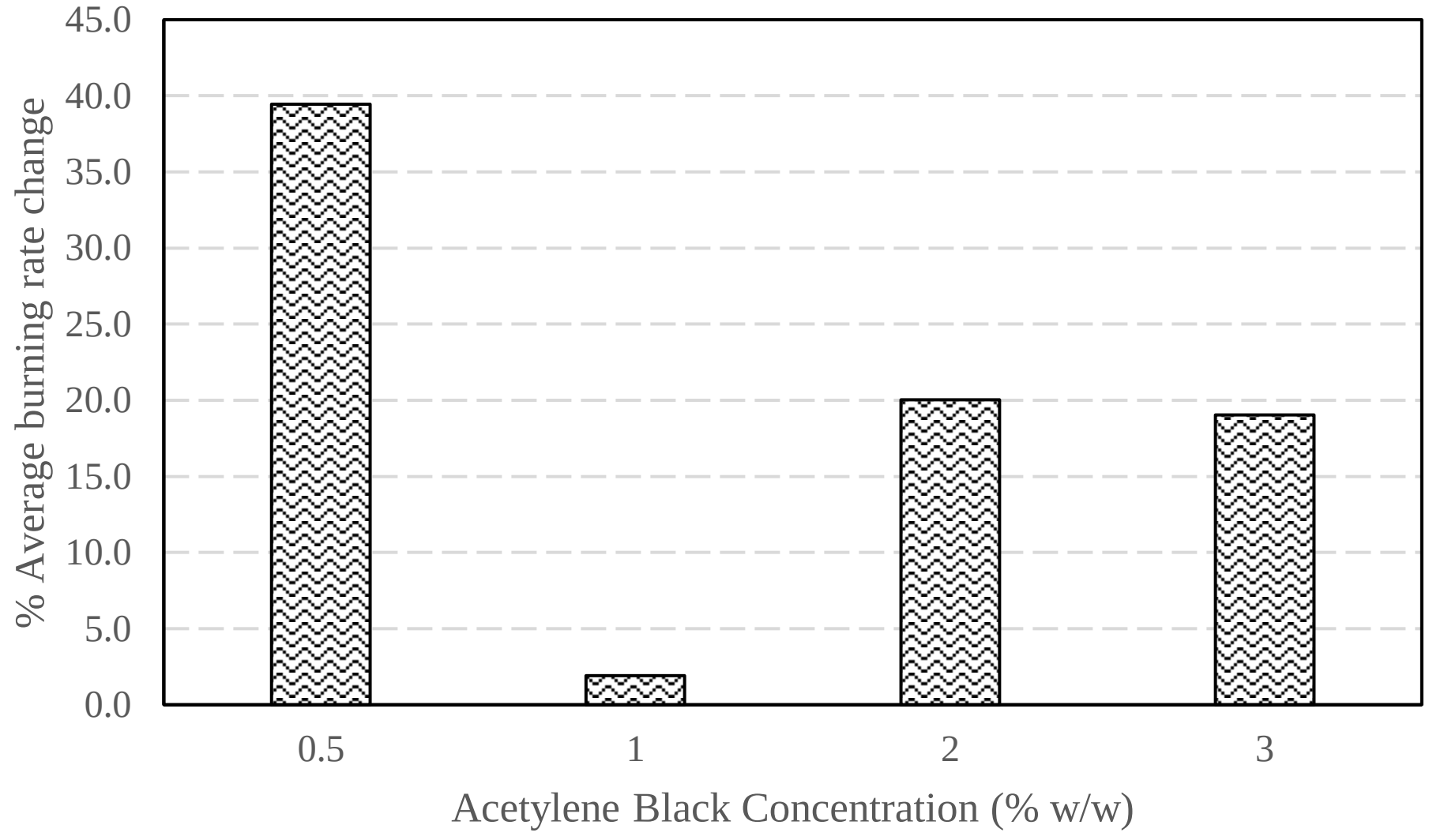

(b)

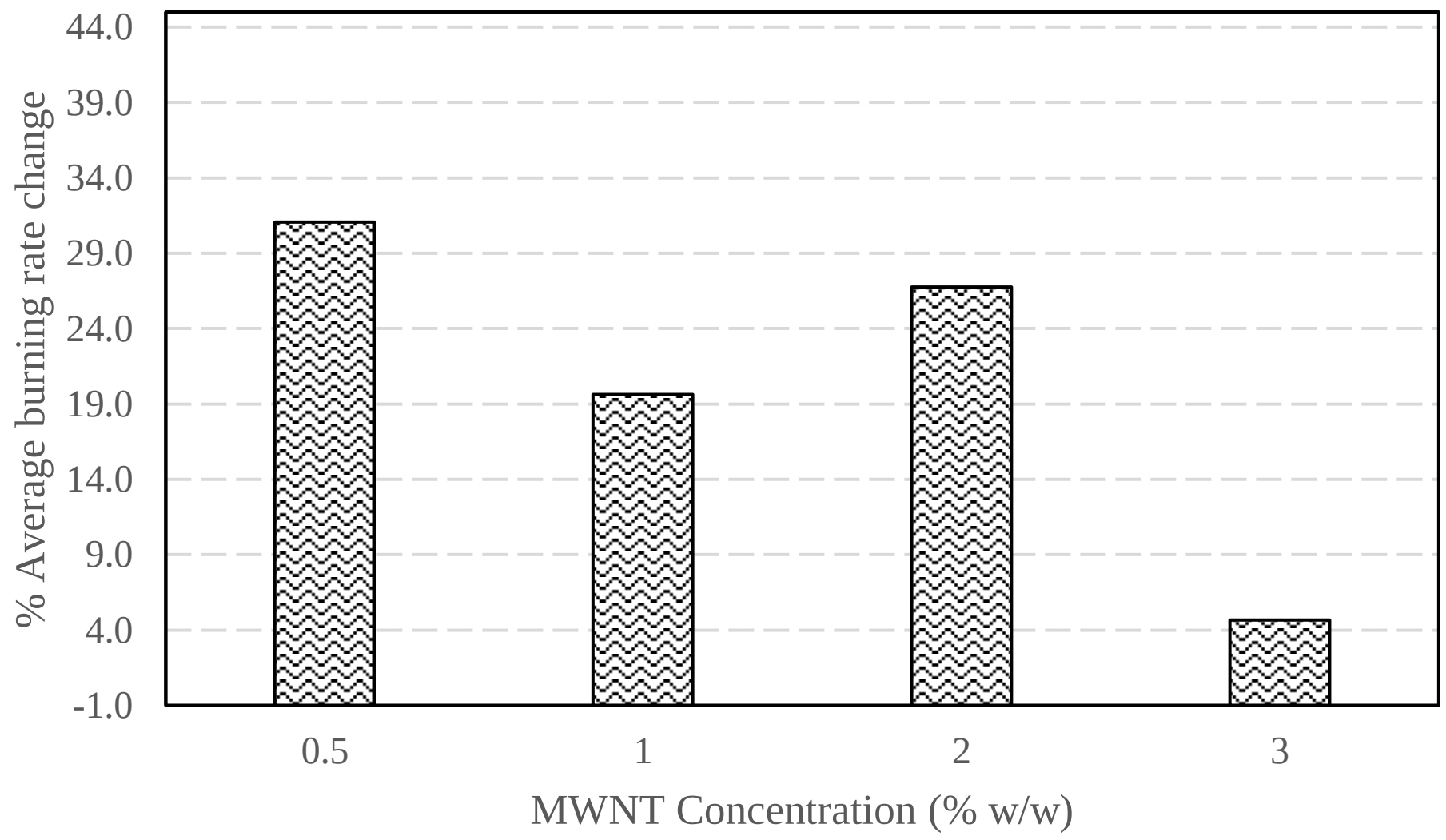

**Figure 6.** Change in average combustion rate of Bakken crude oil at different particle loadings of (a) AB and (b) MWNT.

*3.2 Ignition delay time*

Ignition delay is a parameter governed by many physical processes. In this experiment, the droplet is surrounded by hot coils and continuously heated until it starts to release enough vapor from its surface to catch fire and release heat in a flame. The flame in turn heats the droplet as the hot coils withdraw and keeps the combustion process going. The delay between the coils starting to heat the droplet and the droplet showing a visible flame is termed *ignition delay*. It depends mainly on how fast the droplet heats up (heat conductivity) and how fast it releases vapor (vapor pressure). The lesser the heat conductivity, the more the ignition delay, and the lesser the vapor pressure, the more the ignition delay.

(a)

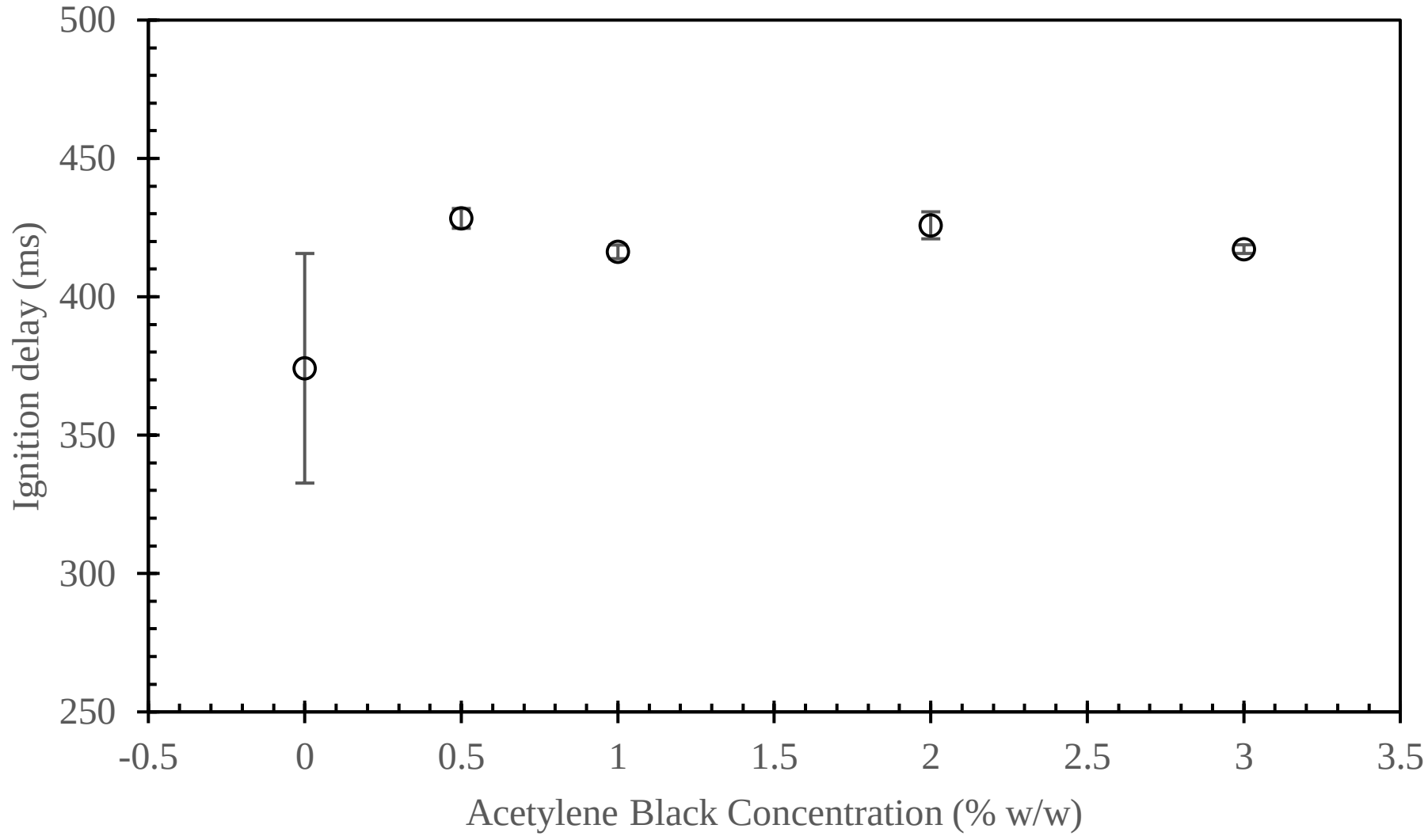

(b)

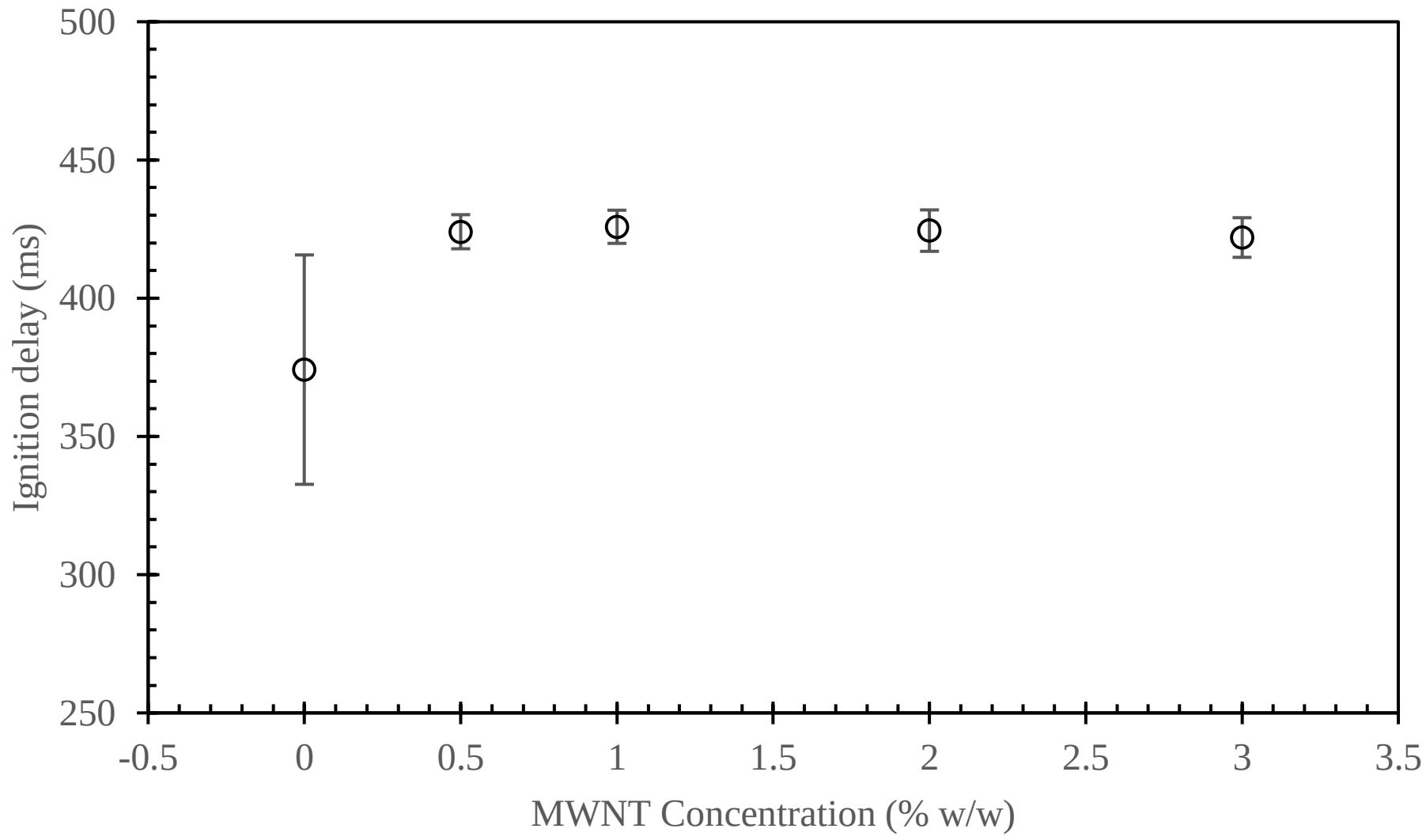

**Figure 7.** Comparison of the ignition delay times of Bakken crude with different mass concentrations of (a) AB and (b) MWNT. Each data point represents an average of six repetitions, and the error bars show the corresponding standard deviation.

Both AB [18] and MWNT [15] show an increased heat conductivity in hydrocarbon-based liquid fuels. Section 3.4 shows that they also decrease vapor pressure. **Figure 7** presents a comparison of the average ignition delay of pure Bakken crude with Bakken crude at different nanomaterial loadings. Six experiments were conducted to determine the average ignition delay for all fuels. The error bars show the standard deviations. In all cases, there is an increase in the ignition delay time for Bakken crude as AB and MWNT are added. Note that the error bar for pure Bakken crude is larger than any of its colloidal suspensions. This can be attributed to the presence of nanomaterials in the liquid bulk, which forms a highly porous and interconnected aggregate inside the burning droplet and causes regularization for the diffusion and evaporation processes at the droplet surface.

(a)

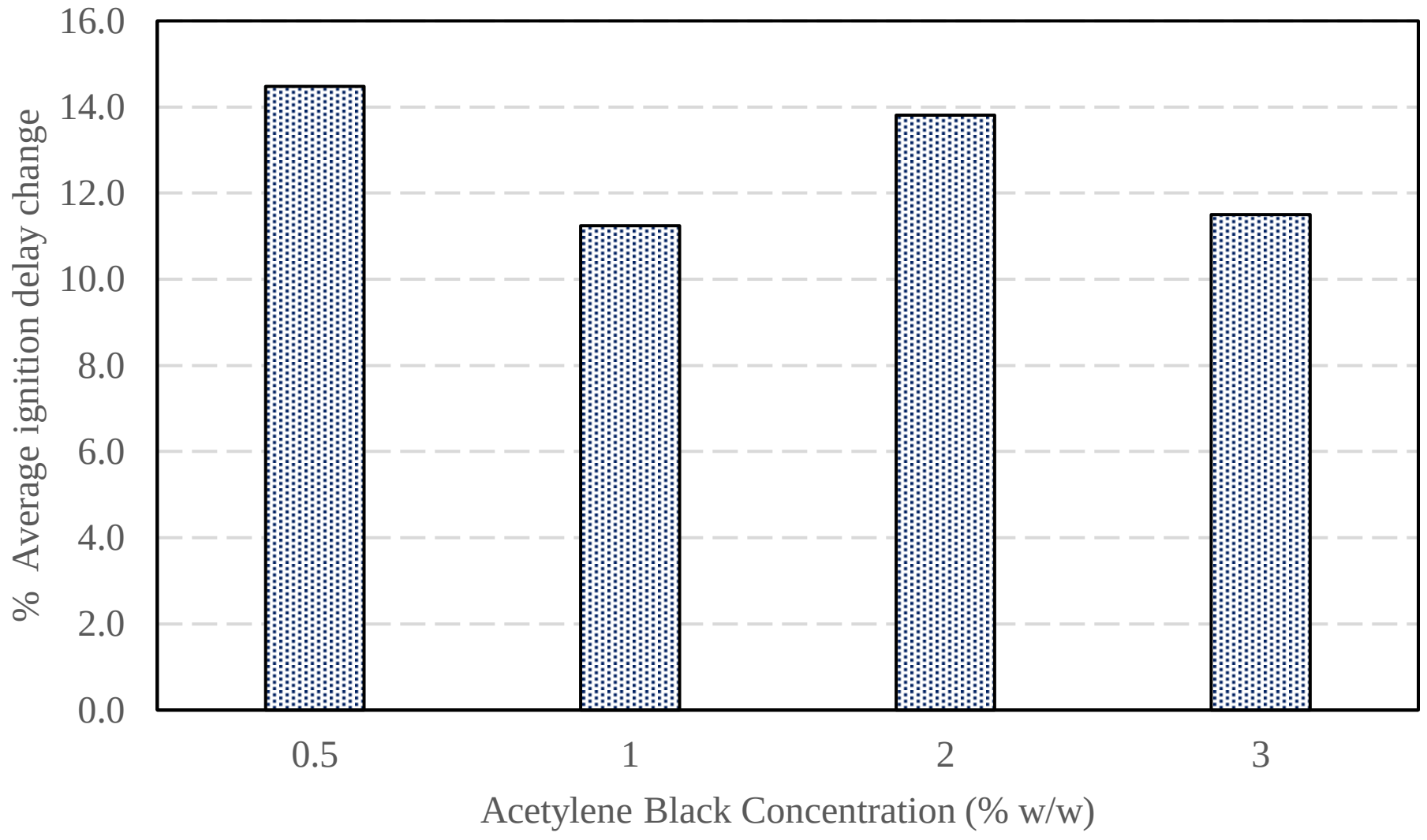

(b)

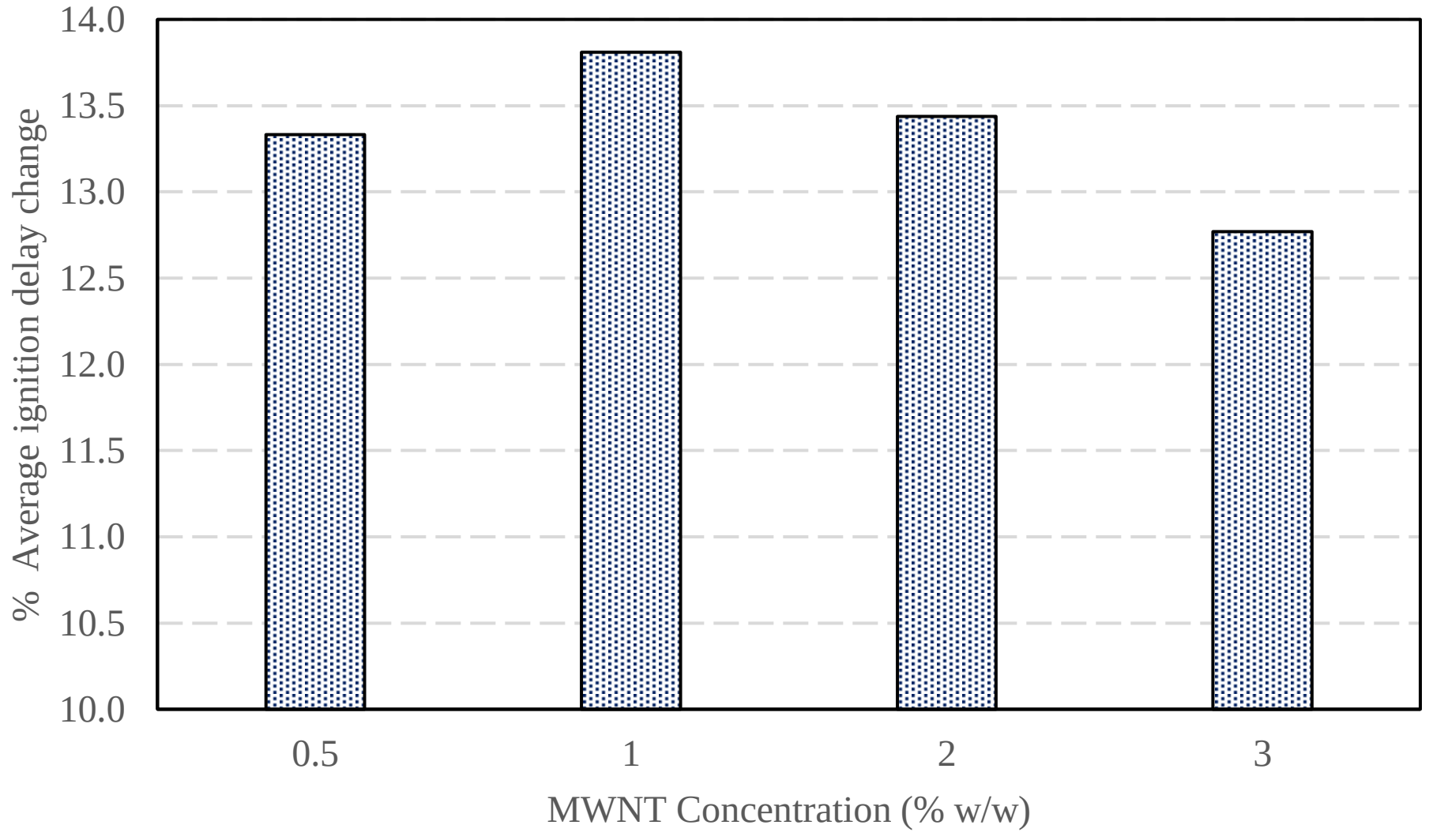

**Figure 8.** Change in average ignition delay time for Bakken crude oil at different particle loadings of (a) AB and (b) MWNT. The largest changes are seen at 0.5% w/w particle loadings for both types of nanoparticles.

**Figure 8** shows the change in average ignition delay for Bakken crude at different particle loadings of AB and MWNT. The largest increase in ignition delay occurs at 0.5% w/w particle loading for AB and 1% w/w particle loading for MWNT. This points to a substantial decrease in vapor pressure without much corresponding increase in heat conductivity at these particle loadings. For AB, the ignition delay increase is smaller at 1% particle loading, meaning the decrease in vapor pressure is counteracted by the increase in heat conductivity. The increase in heat conductivity is more pronounced at 2% AB particle loading with an increase in ignition delay than at 1% AB. However, for both AB an MWNT, the ignition delay is lower at 3% particle loading than at 2%, which points to a greater decrease in vapor pressure than can be counteracted by an increase in heat conductivity.

*3.3 Total combustion time*

**Figure 9** shows the average total combustion time comparison of Bakken crude at different nanomaterial loadings. Six experiments were conducted to determine the average ignition delay for all fuels. The error bars show the standard deviations. It is observed that acetylene black has a significant effect on the total combustion time of Bakken crude, but MWNT does not produce any significant effect of the same magnitude.

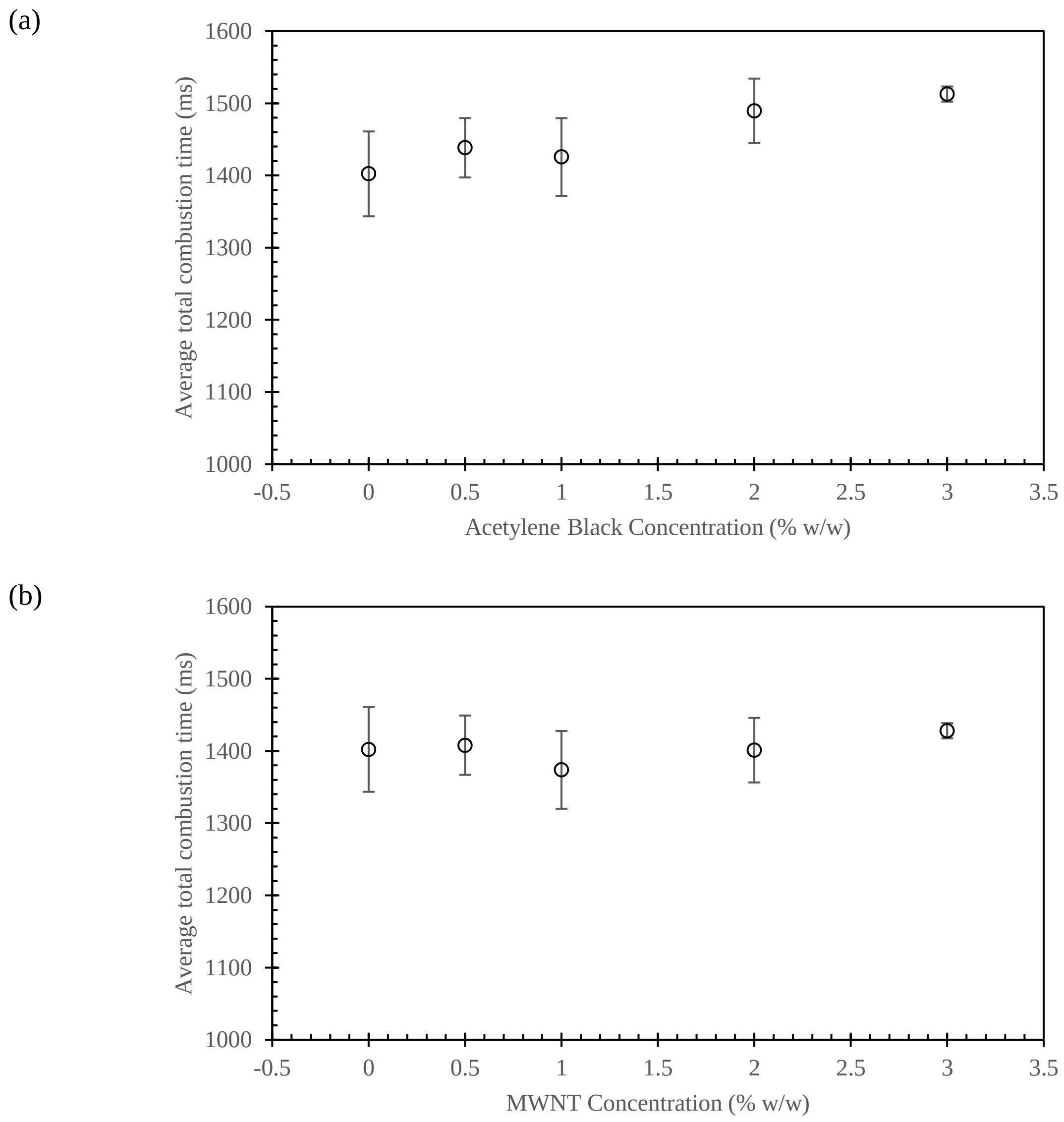

**Figure 9.** Comparison of the total combustion times of Bakken crude with different mass concentrations of (a) AB and (b) MWNT. Each data point represents an average of five repetitions, and the error bars show the corresponding standard deviation.

**Figure 10** shows the change in average total combustion time for Bakken crude at different particle loadings of AB and MWNT. The general trend is an increase in the average total combustion time for Bakken crude as AB particle loading is increased. This is counterintuitive, as there is an increase in combustion rate for Bakken crude as AB is added (Section 3.1). The explanation lies in the decrease in microexplosion intensity for liquid fuels as AB is added, which has previously been reported in the literature [18]. This leads to less liquid fuel being lost during the droplet fragmentation that occurs during a microexplosion event, which results in a greater total combustion time. The maximum increase is 7.9% at 3% AB particle loading. Since MWNT does not have the same effect as AB on controlling microexplosions, the change in total combustion time is marginal for Bakken crude loaded with MWNT: the maximum increase is 1.8% at 3% particle loading, and the maximum decrease is 2% at 1% particle loading.

(a)

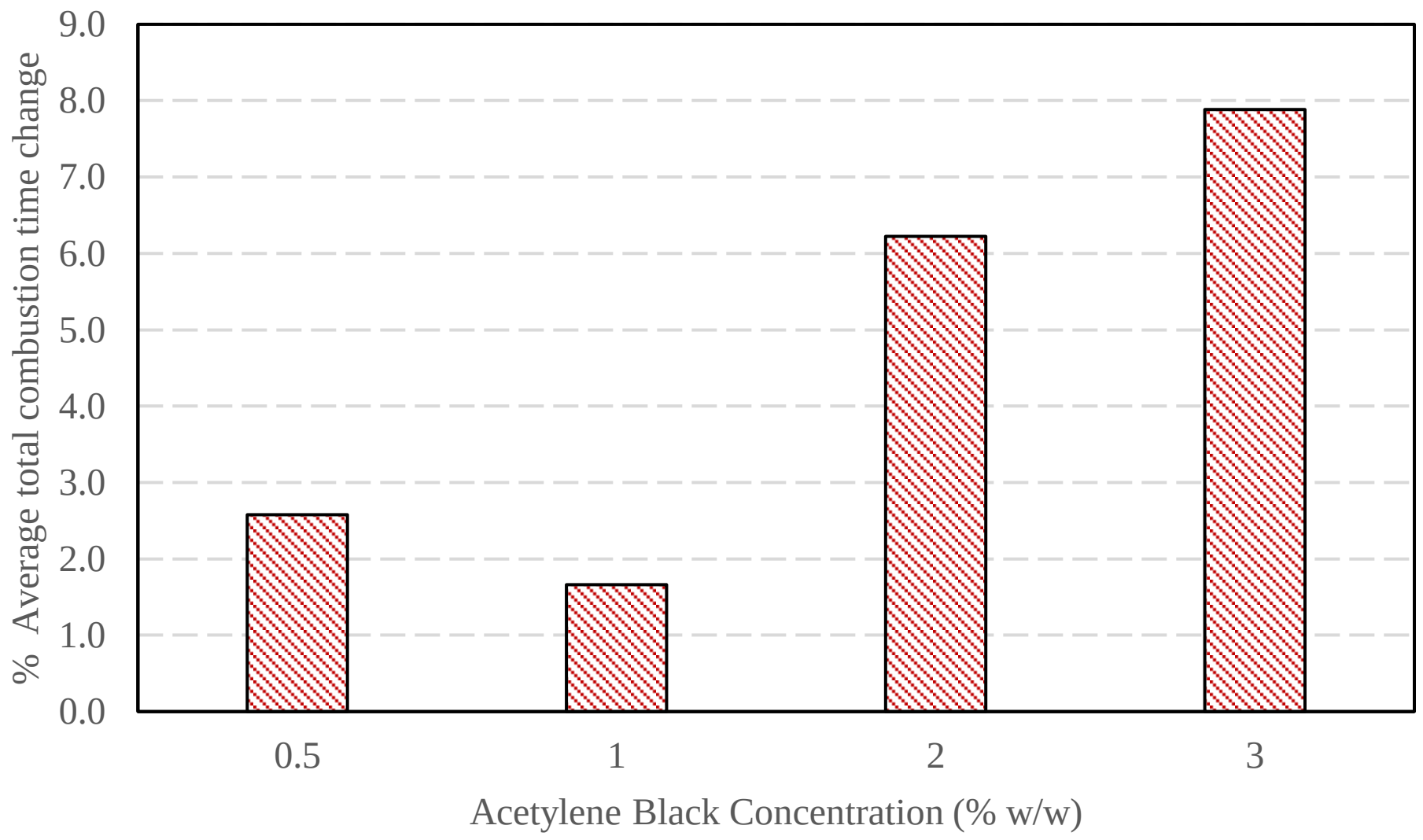

(b)

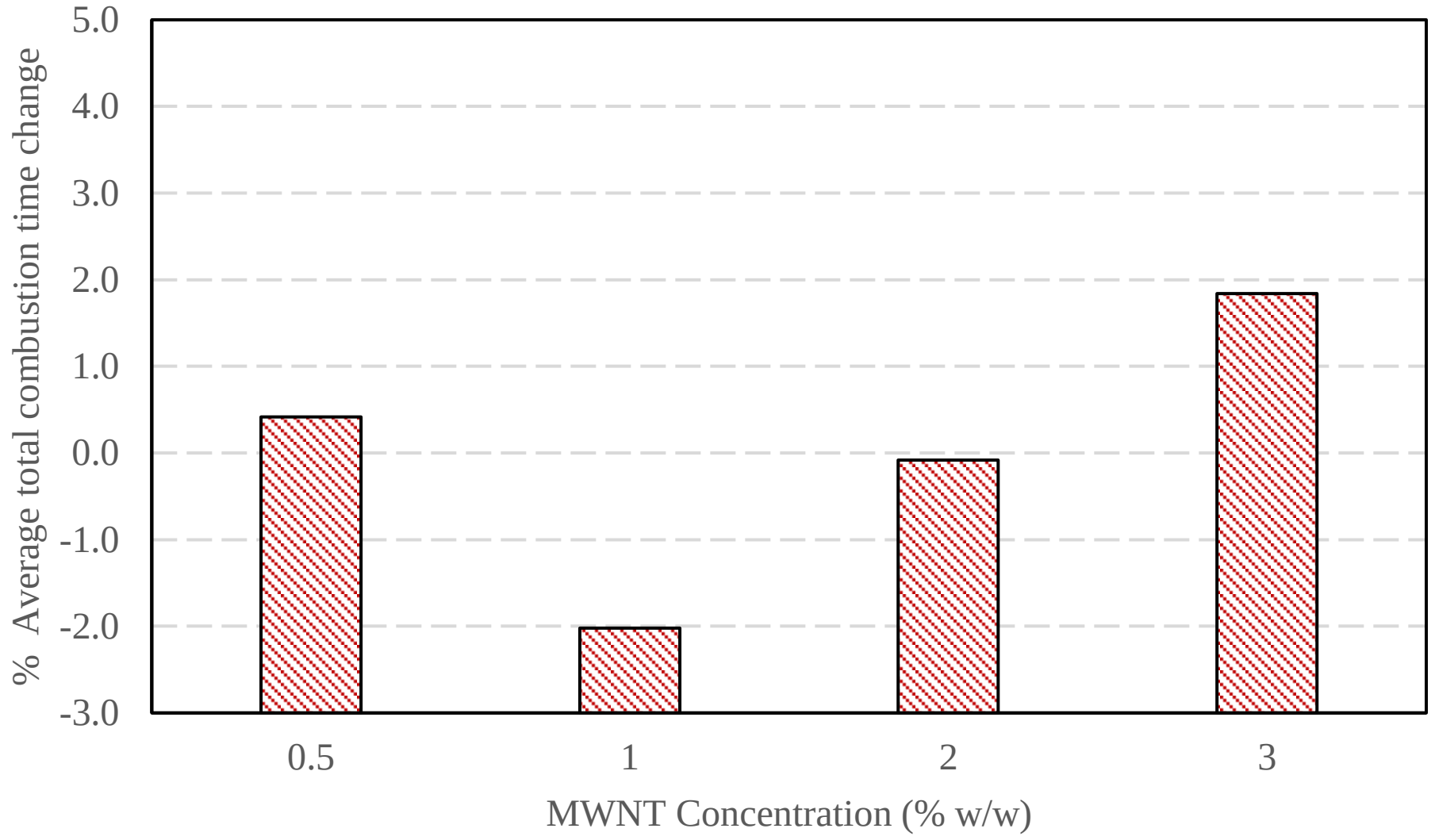

**Figure 10.** Change in average total combustion time for Bakken crude oil at different particle loadings of (a) AB and (b) MWNT.

### *3.4 Flame stand-off ratio*

Flame stand-off ratio (FSR), defined as the ratio of flame diameter $d_f$ to droplet diameter $d_d$, is an important parameter when comparing the combustion processes of different fuels. The regular experimentation process requires a bright backlight to obtain a sharp image of the droplet, which drowns out the flame. Therefore, this experiment was performed in low-light conditions with high exposure settings and the backlight at low power. This allowed a sharp capture of both the droplet and the flame with the CCD camera. **Figure 11** shows a low-light image of a burning Bakken + 0.5% AB fuel droplet, where the flame structure is visible.

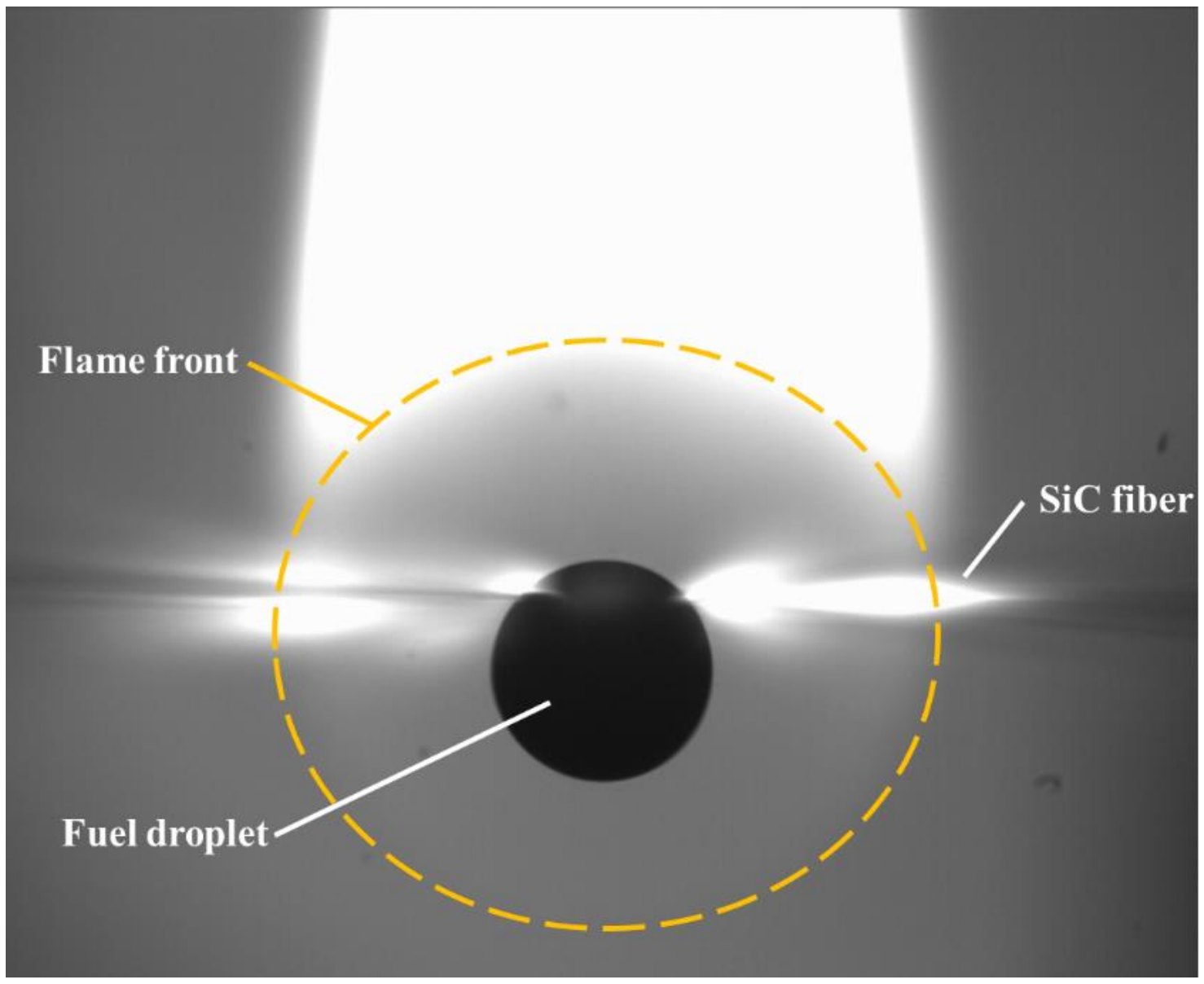

**Figure 11.** Low-light image of Bakken crude oil with 0.5% w/w AB particle loading showing the burning fuel droplet and its flame structure.

The FSR allows for a better understanding of the surface phenomena. There are two competing phenomena that dictate the size of flame diameter $d_f$ compared to droplet diameter $d_d$: thermophoretic flux, which pushes the flame towards the droplet due to the temperature gradient, and the Stefan flux, which pushes the flame away from the droplet because of fuel evaporation from the droplet. Therefore, a fuel with a lower evaporation rate (and lower vapor pressure) will have a lower Stefan flux and, consequently, a lower $d_f$, leading to a higher FSR. **Figure 12** shows the FSR comparison between pure Bakken crude and its colloidal suspensions at different particle loadings. Note that the FSR for the crude with nanomaterial loading is generally lower than for pure crude, meaning that the nanomaterial present at the droplet surface hinders liquid fuel evaporation, which leads to a lower Stefan flux. Therefore, the addition of nanomaterials to a liquid fuel causes its vapor pressure to drop. This also explains the general increase in ignition delay of Bakken crude as it is loaded with nanomaterials: it takes longer for the critical amount of vapor to accumulate around the droplet to start a self-sustaining combustion regime. This behavior may seem at odds with the higher observed combustion rate with nanomaterial addition. It can be explained by the thermal decomposition of fuel constituents due to disproportionately increased droplet temperature when nanomaterials are added. Thermal decomposition leads to long-chain components breaking down into lighter components, which counter the effect of decreased vapor pressure by increasing flame speed and increasing combustion rates, but not enough to overcome the decreased Stefan flux, which leads to a lower FSR for droplets with added nanomaterials.

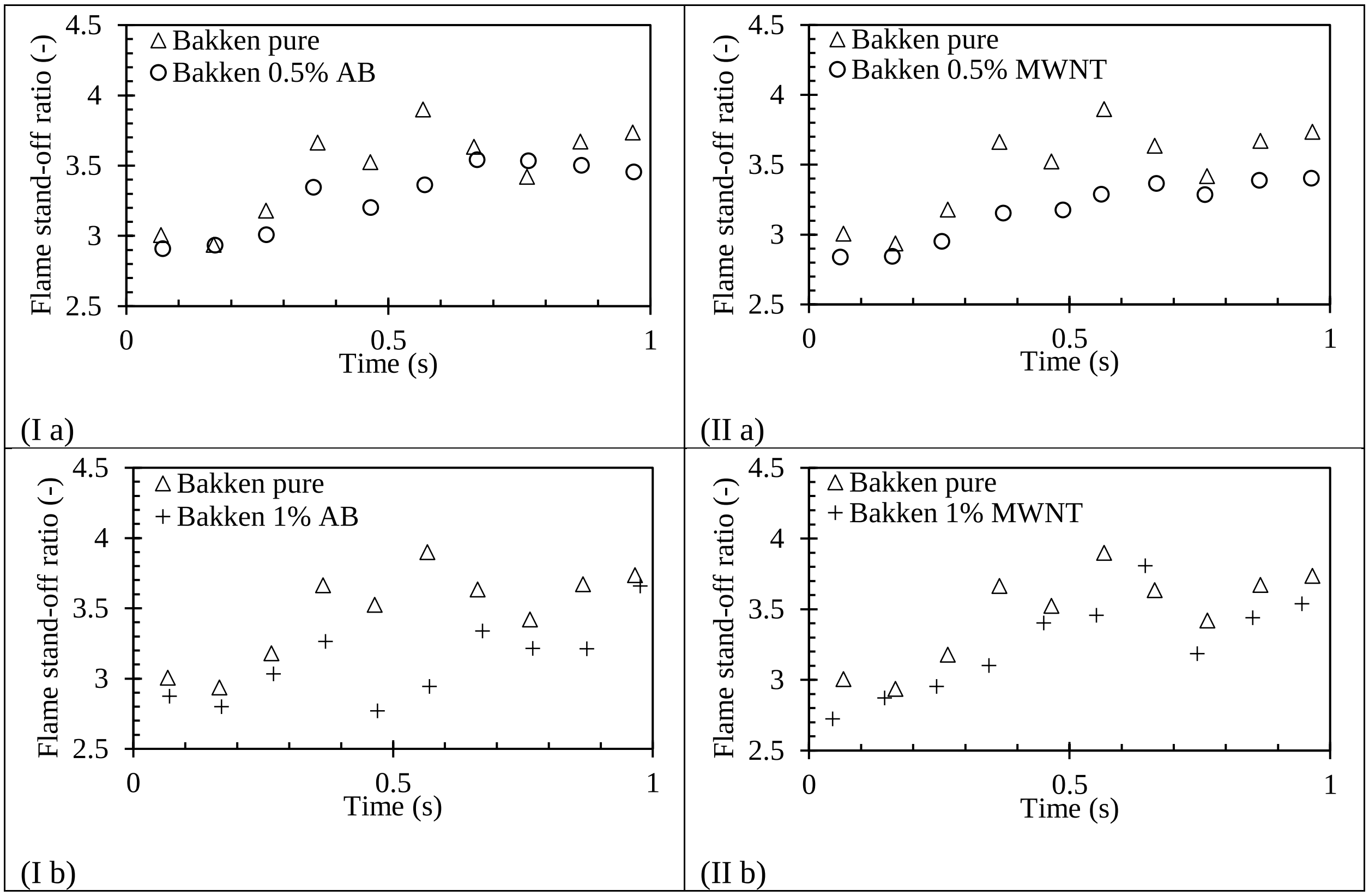

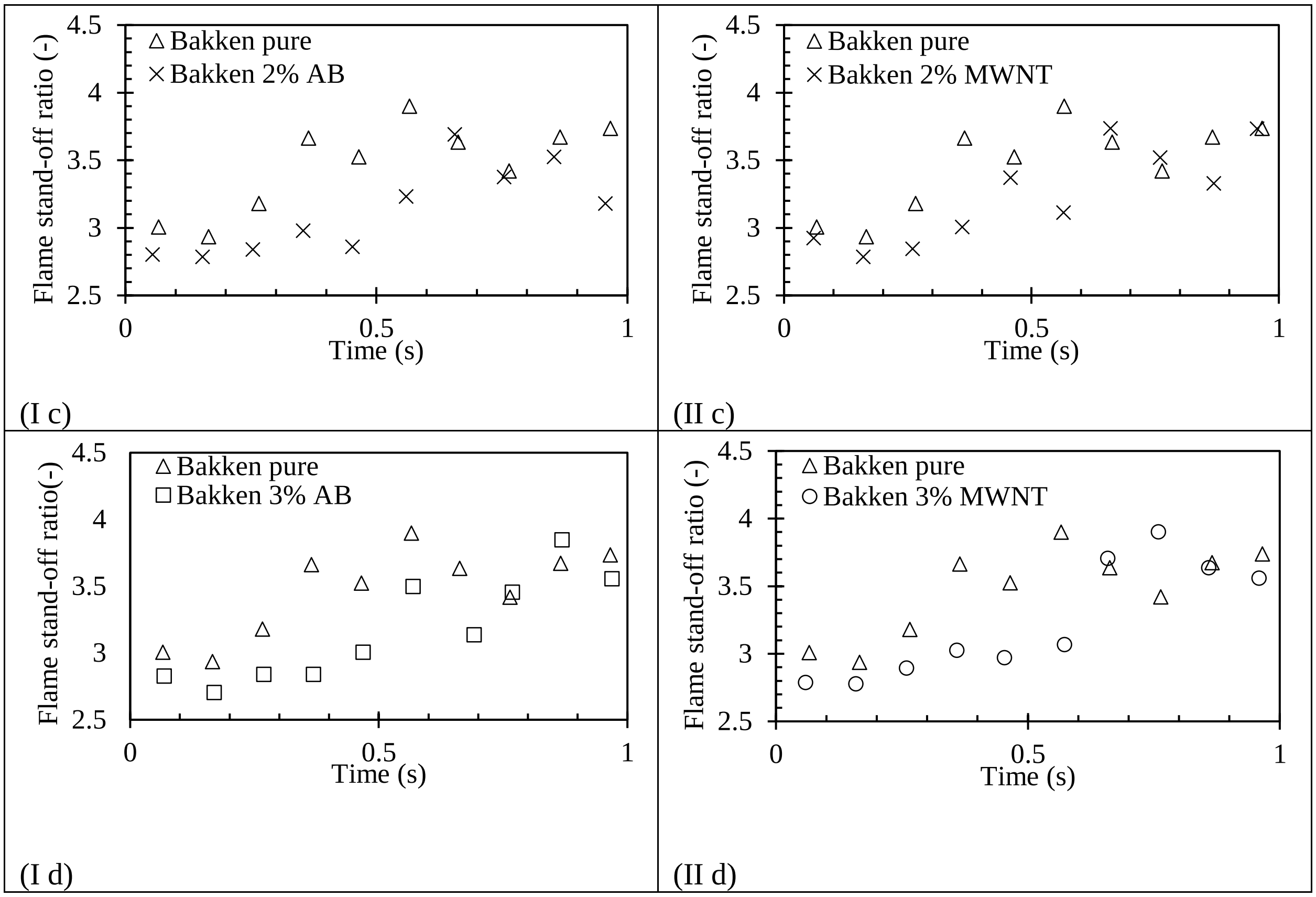

**Figure 12.** Comparison of flame stand-off ratios (FSR) for Bakken crude oil at (a) 0.5%, (b) 1%, (c) 2%, and (d) 3% w/w particle loadings of AB (I) and MNWT (II).

## 4. Conclusions

An experimental investigation of the combustion and flame properties of Bakken crude and its colloidal suspensions with acetylene black (AB) and multi-walled carbon nanotubes (MWNT) at 0.5%, 1%, 2%, and 3% w/w particle loadings was carried out. Sub-millimeter spherical droplets of the fuel of interest were burned to completion, with the process being recorded by CCD and CMOS cameras. The resulting images were post-processed with ImageJ and MATLAB to obtain various parameters such as burning rate, ignition delay, total combustion time, and flame stand-off ratio. Significant changes in all Bakken crude combustion parameters were observed when nanomaterials were added to the crude.

- It was found that the addition of AB and MWNT caused an increase in the bulk heat conductivity and radiation absorption of Bakken crude, but a decrease in its vapor pressure.
- Very low amounts of nanomaterials (0.5% w/w) are required to achieve a significant increase in combustion rates for crude.
- Due to increased heat conductivity and radiation absorption, a maximum of 39.5% and 31.1% combustion rate enhancement was observed at particle loading of 0.5% w/w AB nanoparticles and 0.5% w/w MWNT, respectively.
- Due to decreased vapor pressure, a maximum of 14.5% and 13.8% average ignition delay increase was noted at particle loadings of 0.5% w/w AB and 1.0% w/w MWNT, respectively.

- Furthermore, a maximum of 9.2% and 1.8% average total combustion time increase was noted at particle loadings of 3% w/w AB and 3% w/w MWNT, respectively.

Various combustion properties, such as combustion rate, are expected to be used in future work to validate the numerical modeling for multi-component, multi-phase fuels with nano-additives. It is expected that this work will stimulate more interest in the addition of AB and MWNT nanoparticles to oil spills to increase ISB effectiveness.

**Acknowledgements**

This research is funded, in part, by the Mid-America Transportation Center via a grant from the U.S. Department of Transportation's University Transportation Centers Program, and this support is gratefully acknowledged. The USDOT UTC grant number for MATC is 69A3551747107. The authors would also like to acknowledge use of the University of Iowa High Resolution Mass Spectrometry Facility (HRMSF). We would especially like to thank Prof. Lynn M. Teesch and Mr. Vic R. Parcell for their help and support with the GC-MS data. The contents reflect the views of the authors, who are responsible for the facts and the accuracy of the information presented herein, and are not necessarily representative of the sponsoring agencies, corporations, or persons.

**Appendix A. Gas Chromatography – Mass Spectrometry (GC-MS) data for Bakken crude oil**

GC-MS data was generated at the High Resolution Mass Spectrometry Facility (HRMSF) located at the University of Iowa Department of Chemistry (**Figure** ). The column used was a 30 m DB-5MS, 0.25mm diameter and 0.25 micrometer film thickness. The temperature ramp started at 50 ºC and held for 1 min. It was then increased at 10 ºC/min until 320 ºC and then held for 5 min. The numbers on top of the peaks in the chromatogram are retention time, area, and response height. The results reveal a liquid rich in low- and medium-boiling components.

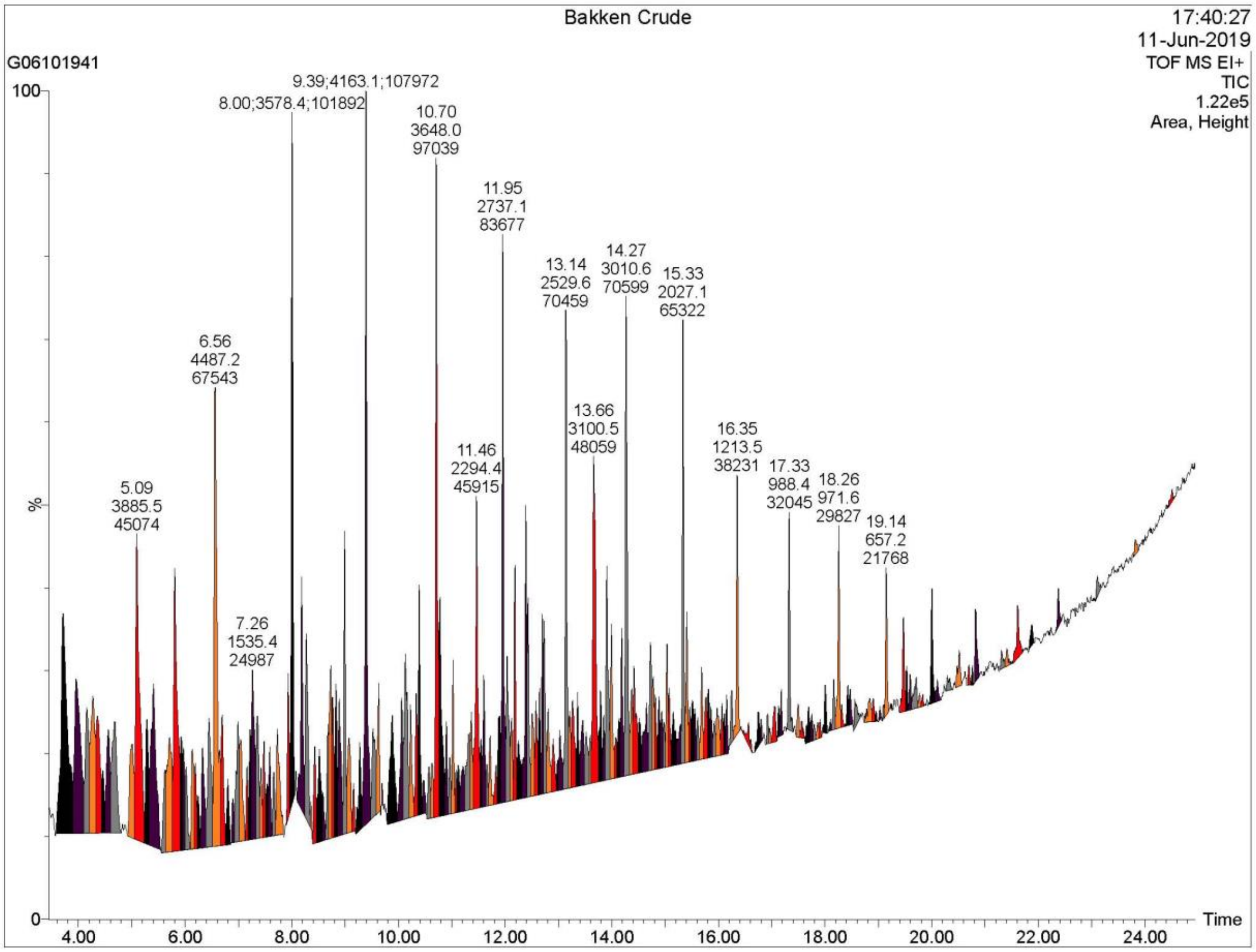

**Figure A1.** Gas Chromatography – Mass Spectrometry (GC-MS) data for Bakken crude oil.